\documentclass{article}
\usepackage{amsfonts,amssymb,amsmath}
\usepackage{graphicx}
\newcommand{\R}{\mathbb{R}}

\begin{document}
\title{The Remarkable Simplicity of Very High Dimensional Data: Application
of Model-Based Clustering}

\author{Fionn Murtagh\thanks{Department of Computer Science,
Royal Holloway, University of London, Egham TW20 0EX, England.
{\tt fmurtagh@acm.org}}
}
\markboth{Very High Dimensional Data: Model-Based Clustering}
{Very High Dimensional Data: Model-Based Clustering}
\maketitle

\begin{abstract}
An ultrametric topology formalizes the notion of hierarchical structure.
An ultrametric embedding, referred to here as ultrametricity, is 
implied by a hierarchical embedding.  Such hierarchical 
structure can be global in the data set, or local.  
By quantifying extent or degree of ultrametricity in a data set, we show
that ultrametricity becomes pervasive as dimensionality and/or
spatial sparsity increases.  This leads us to assert that very high
dimensional data are of simple structure.  We exemplify this finding through
a range of simulated data cases.  We discuss also application to very 
high frequency time series segmentation and modeling.
\end{abstract}



\section{Introduction}

The lessons we will draw from this work are as follows. 

\begin{itemize}
\item Very high dimensional spaces are of very simple structure.
\item It becomes easier to find clusters in high dimensions.
\item The simple high dimensional structure is hierarchical.
\item Ease of handling high dimensional data, e.g.\ reading off
clusters,  emulates the human perception system which similarly processes
data with no evident latency.
\end{itemize}

There is a burgeoning crisis in high dimensional data analysis, and
many current approaches lack convincing performance guarantees.
In Hinneburg, Aggarwal and Keim (2000), 
attention is focused on ``relevant'' dimensions only,
while Aggarwal, Hinneburg and Keim (2001) 
(cf.\ the revealing titles in the case of both
of these citations) state:
``Recent research results show that in high dimensional space,
the concept of proximity, distance or nearest neighbor may not even
be qualitatively meaningful.''  The last-mentioned work investigates $L_p$ norms
including for fractional values of $p$.

In Breuel (2007), the focus is the
$\epsilon$-approximate nearest neighbor defined as follows: if the
nearest neighbor point $y$ to some query point, $x$,
has distance $d(x,y)$, then any vector $y'$ such that $d(y', x)
\leq (1 + \epsilon) d(x,y)$ is an $\epsilon$-approximate nearest
neighbor of $x$.  Then Breuel (2007) points out: ``... the relationship between
approximation and 'cost' of a solution need not be linear.  For
example, the cost of picking an $\epsilon$-approximate nearest
neighbor could be proportional not to the difference of distances
between the optimal answer and the approximation, but to the volume of
the shell between the two, that is, as $(1 + \epsilon)^{m-1}$, where
$m$ is the dimension of the space.''  Various issues immediately
ensue: (i) ``for large dimensions, even small values of $\epsilon$
include the entire database as $\epsilon$-approximate neighbors''; (ii)
``analyzing the worst-case asymptotic complexity of
$\epsilon$-approximate algorithms is meaningless''; (iii) for ``large enough
dimensions, a randomly chosen point becomes an
$\epsilon$-approximate nearest neighbor with high probability''; (iv)
``the implicit assumption that a close approximation leads to only a
small increase in the cost of a solution is not justifiable in the
context of nearest neighbors''; and (v) an $\epsilon$-approximate nearest
neighbor algorithm is not necessarily useless in practice but such an
algorithm has
``neither useful meaning asymptotically (as the dimension grows),
nor does it make useful predictions about its behavior on practical
problems''.

In this article, we propose an  approach which we consider appropriate
for high dimensional data analysis, based on clustering and proximity
searching.

\subsection{The Ultrametricity Perspective and Overview of this Work}

The morphology or inherent shape and form of an object is important.  In 
data analysis, the inherent form and structure of data clouds are important.
So the embedding topology, with which the data clouds are studied, can 
be crucial.  
Quite a few models of data form and structure are used in data analysis.  
One of them is a hierarchically embedded set of clusters, -- a 
hierarchy.  It is traditional (since at least the 1960s) to impose such a
form on data, and if useful to assess the goodness of fit.  Rather than
fitting a hierarchical structure to data (e.g., Rohlf and Fisher, 1968), 
our recent work
has taken a different orientation: we seek to find (partial or global) 
inherent hierarchical structure in data.  As we will describe in this 
article, there are interesting findings that result from this, and some
very interesting perspectives are opened up for data analysis and, potentially,
perspectives also on the physics (or causal or generative mechanisms) 
underlying the data.  

A formal definition of hierarchical structure is provided by ultrametric
topology (in turn, related closely  to p-adic number theory).  We will
return to this in section \ref{sect23} below.  First, though, we will 
summarize some of our findings.  

Ultrametricity is a pervasive property of observational data.  It 
arises as a limit case when data dimensionality or sparsity grows.  More
strictly such a limit case is a regular lattice structure and
ultrametricity is one possible representation for it.  Notwithstanding 
alternative representations, ultrametricity offers
 computational efficiency (related to tree depth/height being logarithmic 
in number of terminal nodes), linkage with dynamical or related 
functional properties (phylogenetic interpretation), and 
processing tools based on well known p-adic or ultrametric theory (examples: 
deriving a partition, or applying an ultrametric wavelet transform).  In 
Khrennikov (1997) and other works, Khrennikov has pointed to the importance
of ultrametric topological analysis.  

Local ultrametricity is also of importance.  
This can be used for forensic data 
exploration (fingerprinting data sets):  see Murtagh (2005, 2007); 
and to expedite search and discovery in 
information spaces: see 
Ch\'avez, Navarro, Baeza-Yates and Marroqu\'{\i}n (2001) as discussed by us in 
Murtagh (2004, 2006) and Murtagh, Downs and Contreras (2007). 

In section \ref{sect23} we show how extent of ultrametricity is measured.  
Section \ref{sect3} presents our main results on the remarkable properties
of very high dimensional, or very sparse, spaces.  As dimensionality 
or sparsity grow, so does the inherent hierarchical nature of the data 
in the space.  In section \ref{appsect4} we then discuss application to 
very high frequency time series modeling.  

\subsection{Review of Recent Asymptotic Statistical Findings}

We can characterize clustering algorithms in terms of number of 
observables, $n$, and number of attributes, $m$, where by ``large'' is meant
thousands upwards: (i) large $n$, small $m$; as is fairly standard in 
astronomy;  (ii) large $n$, large $m$; as is fairly typically the case 
in information retrieval; and (iii) small $n$, large $m$; as is often the 
case in bioinformatics, and textual forensics.  It is case (iii)  which is 
of most interest to us here.  However our results also accommodate case (ii). 

In Hall, Marron and Neeman (2005), 
it was shown that ``under 
some conditions on underlying distributions, as the dimension tends to infinity
with a fixed sample size, the $n$ data vectors form a regular $n$-simplex in 
$\R^m$'' (Ahn, Marron, Muller and Chi, 2007).   These authors term the small $n$, 
large $m$ case
``HDLSS, high dimension, low sample size''.   In Ahn and Marron (2005), 
some other 
unrelated work is cited, where the ratio of $m/n$ tends to a constant.  As with 
these authors, our goal is to study the case of letting ``$m$ tend to infinity,
fixing $n$'' (Ahn and Marron, 2005).  Hall et al.\ (2005) discuss previous 
asymptotics work in the statistical literature.  
Our focus is not on a simplex but rather 
on a hierarchy (even if trivial) in order to study implications for data 
analysis. 

For ``the asymptotic  geometric
representation of HDLSS data'', it is shown in the work of Ahn and collaborators 
that ``when 
$m >> n$, under a mild assumption, the pairwise distances between each pair 
of data points are approximately identical so that the data points form a 
regular $n$-simplex.  In a binary classification setting, the training data 
from each class becomes two simplices ... any reasonable classification method 
will find the same [discriminant result] when $m$ becomes very large.'' 
(Ahn et al., 2007).
This is a very exciting for the discriminant analysis case pursued by
Ahn et al., 2007; Ahn and Marron, 2005; Hall et al., 2005) 
(naive Bayes, SVM or support vector machine, Fisher's 
linear discriminant, and the simplex structure in very high dimensions leading
to the ``direction of maximal data piling'').  In this paper, our interest 
is in clustering, or unsupervised classification.  

The mild condition for simplex structure formation, as $m \longrightarrow \infty$
is that directionality of the Gaussian cloud is ``diffuse'', defined in 
terms of eigenvalues:

$$ \sum^m_j \lambda_j^2 / \left( \sum^m_j \lambda_j \right)^2   \longrightarrow
0 \mbox{  as  }  m \longrightarrow \infty $$

Then it is shown that the covariance matrix approaches a constant times 
the identity matrix.  

In Donoho and Tanner (2005), the Gaussian case is also focused on.  For a Gaussian
cloud, ``not only are the points on the convex hull, but all reasonable-sized
subsets span faces of the convex hull''.  Intuitively, if all points fly
apart from one another as dimensionality grows, then (i) each point is a vertex
of the convex hull of the cloud of points; (ii) each pair of points generates
an edge of the convex hull; and (iii) sets of points form a
regional face of the convex hull.  These properties are proven by
Donoho and Tanner (2005) who conclude: ``This is wildly different than the behavior
that would be expected by traditional low-dimensional thinking.''

We may ask why  we (in this work) lay importance on the fact that the 
high dimensional simplex
additionally defines an ultrametric topological embedding.
An ultrametric topology requires (as will be described in sections to follow)
 any triangle to be either
(i) equilateral, or (ii) isosceles with small base.
The equilateral case corresponds fine with the simplex structure.
But it is useful to us to hang on to the isosceles with small base case,
too, for inter-cluster relationships.
We will look later at examples to support this viewpoint.  

\section{Quantifying Degree of Ultrametricity}
\label{sect23}

Summarizing a full description in Murtagh (2004) we explored two
measures quantifying how ultrametric a data set is, -- Lerman's and a new
approach based on triangle invariance (respectively, 
the second and third approaches described in this section).

The triangular inequality holds for a metric space: $d(x,z) \leq 
d(x,y) + d(y,z)$ for any triplet 
of points $x,y,z$.  In addition the properties 
of symmetry and positive definiteness are respected.  The ``strong 
triangular inequality'' or ultrametric inequality is: $d(x,z) \leq 
\mbox{ max } \{ d(x,y), d(y,z) \}$ for any triplet $x,y,z$.  An
ultrametric
space implies respect for a range of stringent properties.  For
example, 
the triangle formed by any triplet is necessarily isosceles, with the
two
large sides equal; or is equilateral.

\begin{itemize}
\item 
Firstly, Rammal, Toulouse and Virasoro (1986) used discrepancy between 
each pairwise 
distance and the corresponding subdominant ultrametric.  Now, the
subdominant ultrametric is also known as the ultrametric distance
resulting from the single linkage agglomerative hierarchical
clustering method.   Closely related graph structures include the
minimal spanning tree, and graph (connected) components.  
While the subdominant provides a good fit to the
given distance (or indeed dissimilarity), it suffers from the
``friends of friends'' or chaining effect.  

\item
Secondly, Lerman (1981) developed a measure of ultrametricity,
termed H-classifiability,  using
ranks of all pairwise given distances (or dissimilarities).  The
isosceles (with small base) or equilateral requirements of the
ultrametric inequality impose constraints on the ranks.  The interval between
median and maximum rank of every set of triplets must be empty for 
ultrametricity.  We have used extensively  
Lerman's measure of degree of ultrametricity in a data set.  
Taking ranks provides scale invariance.  
But the limitation of Lerman's approach, 
we find, is that it is not reasonable to
study ranks of real-valued (values in non-negative reals) 
distances defined on a large set of points.

\item
Thirdly, our own measure of extent of ultrametricity (Murtagh, 2004)
can be described algorithmically.  We 
examine triplets of points (exhaustively if possible, or otherwise
through sampling), and determine the three angles formed by the
associated triangle.  We select the smallest angle formed by the triplet
points.  Then we check if the other two remaining angles are
approximately equal.  If they are equal then our triangle is isosceles
with small base, or equilateral (when all triangles are equal).  The
approximation to equality is given by 2 degrees (0.0349 radians).  
Our motivation for
the approximate (``fuzzy'') equality is that it makes our approach
robust and independent of measurement precision.  
\end{itemize}

A supposition for use of our measure of ultrametricity is that we can  
define angles (and hence triangle properties).  This in turn presupposes a
scalar product.  Thus we presuppose a complete 
normed vector space with a scalar product --
as one example, the real part of a Hilbert space -- to provide our needed 
environment.  

Quite a general way to 
embed data, to be analyzed, in a Euclidean space, is to use correspondence 
analysis (Murtagh, 2005).  This explains our interest in using correspondence 
analysis: it provides a convenient and versatile 
way to take input data in many varied formats (e.g., ranks or scores, 
presence/absence,
frequency of occurrence, and many other forms of data) and map them into a 
Euclidean, factor space.

\section{Ultrametricity and Dimensionality}
\label{sect3}

\subsection{Distance Properties in Very Sparse Spaces}
\label{sect22}

Murtagh (2004), and earlier work by Rammal, Angles d'Auriac and Doucot (1985) and
Rammal et al.\ (1986), 
has demonstrated the pervasiveness of ultrametricity, by 
focusing on the fact that  
sparse high-dimensional data tend to be ultrametric.  
In such work it is shown how numbers of points
in our clouds of data points are irrelevant; but what counts is the
ambient spatial dimensionality.  Among cases looked at are statistically 
uniformly
(hence ``unclustered'', or without structure in a certain sense)
distributed points, and statistically 
uniformly distributed hypercube vertices (so the
latter are random 0/1 valued vectors).  Using our ultrametricity
measure, there is a clear tendency to ultrametricity as the spatial
dimensionality (hence spatial sparseness) increases.  



As Hall et al.\ (2005) also show, Gaussian data behave in the same way and a 
demonstration of this is seen in Table \ref{tabunifgauss}.  
To provide an idea of consensus of these
results, the 200,000-dimensional Gaussian was repeated and yielded on 
successive runs values of the ultrametricity measure of: 0.96, 0.98, 0.96.  

\begin{table}
\begin{center}
\begin{tabular}{lllll} \hline
No. points &    Dimen. &  Isosc. &   Equil. &    UM   \\ \hline    
           &           &         &          &         \\
Uniform    &           &         &          &         \\
           &           &         &          &         \\
100        &    20     &    0.10 &    0.03  &    0.13 \\
100        &    200    &    0.16 &    0.20  &    0.36 \\
100        &    2000   &    0.01 &    0.83  &    0.84 \\
100        &    20000  &    0    &    0.94  &    0.94 \\
           &           &         &          &         \\
Hypercube  &           &         &          &         \\
           &           &         &          &         \\
100        &    20     &    0.14 &   0.02   &    0.16 \\
100        &    200    &    0.16 &   0.21   &    0.36 \\
100        &    2000   &    0.01 &   0.86   &    0.87 \\
100        &    20000  &    0    &   0.96   &    0.96 \\
           &           &         &          &         \\
Gaussian   &           &         &          &         \\
           &           &         &          &         \\
100        &    20     &    0.12 &    0.01  &    0.13 \\
100        &    200    &    0.23 &    0.14  &    0.36 \\
100        &    2000   &    0.04 &    0.77  &    0.80 \\
100        &    20000  &    0    &    0.98  &    0.98 \\ \hline
\end{tabular}
\end{center}
\caption{Typical results, based on 300 sampled triangles from triplets of
points.  For uniform, the data are generated on [0, 1]; hypercube vertices
are in $\{ 0, 1\}^m$, 
and for Gaussian, the data are
of mean 0, and variance 1.  Dimen.\ is the ambient
dimensionality.  Isosc.\ is the number of isosceles triangles with 
small base, as a
proportion of all triangles sampled.  Equil.\ is the number of equilateral 
triangles as a proportion of triangles sampled.  UM is the proportion of 
ultrametricity-respecting triangles (= 1 for all ultrametric).}
\label{tabunifgauss}
\end{table}

In the following, we explain why high dimensional and/or sparsely populated
 spaces are ultrametric.  We use the Euclidean distances in the cosine formula
to determine angles.  Note that there is no averaging of distances involved, 
nor distances normalized by dimensionality.  

As dimensionality grows, so too do distances (or indeed
dissimilarities, if
they do not satisfy the triangular inequality).  The least change
possible for dissimilarities to become distances has been formulated
in terms of the smallest additive constant needed, to be added to all
dissimilarities (Torgerson, 1958; Cailliez and Pag\`es, 1976; Cailliez, 
1983; Neuwirth and Reisinger, 1982; and the comprehensive review of 
B\'enass\'eni, Bennani Dosse and Joly, 2007). 
Adding a sufficiently large 
constant to all dissimilarities transforms them into a set of
distances.  Through addition of a larger constant, it follows that
distances become approximately equal, thus verifying a trivial case of
the ultrametric or ``strong triangular'' inequality.  Adding to
dissimilarities or distances may be a direct consequence of increased
dimensionality. 

For a close fit or good approximation,  
the situation is not as simple for taking dissimilarities, or
distances,
into ultrametric distances.  A best fit solution is given by de Soete
(1986)  
(and software is available in Hornik, 2005).
If we want a close fit to the given 
dissimilarities then a good choice would avail either of the maximal 
inferior, or subdominant, ultrametric; or the minimal superior
ultrametric.
Stepwise algorithms for these are commonly known as, respectively,
single
linkage hierarchical clustering; and complete link hierarchical
clustering.  
(See Benz\'ecri, 1979; Lerman, 1981; Murtagh, 1985; and other texts on 
hierarchical clustering.)  

\subsection{No ``Curse of Dimensionality'' in Very High Dimensions}
\label{sect21}

Bellman's (1961)  ``curse of dimensionality'' relates to exponential 
growth of hypervolume as a function of dimensionality.  Problems
become tougher as dimensionality increases.  In particular problems 
related to proximity search in high-dimensional spaces tend to become
intractable.  

In a way, a ``trivial limit'' (Treves, 1997) 
case is reached as dimensionality increases.  This makes high
dimensional 
proximity search very different, and given an appropriate data
structure -- such as a binary hierarchical clustering tree -- we can
find nearest neighbors in worst case $O(1)$ or constant computational
time (Murtagh, 2004).  The proof is simple: the tree data structure
affords a constant number of edge traversals.  

The fact that limit properties are ``trivial'' makes them no less 
interesting to study.  Let us refer to such ``trivial'' properties as
(structural or geometrical) regularity properties (e.g.\ all points
lie on a regular lattice).  

First of all, the symmetries of regular
structures in our data may be of importance.  For example, processing 
of such data can exploit these regularities.  

Secondly, ``islands'' or
clusters in our data, where each ``island'' is of regular structure,
may be of interpretational value.  

Thirdly, and finally, regularity of particular properties does not 
imply regularity of all properties.  So, for example, we may have only 
partial existence of pairwise linkages.  

\begin{figure*}
\includegraphics[width=16cm]{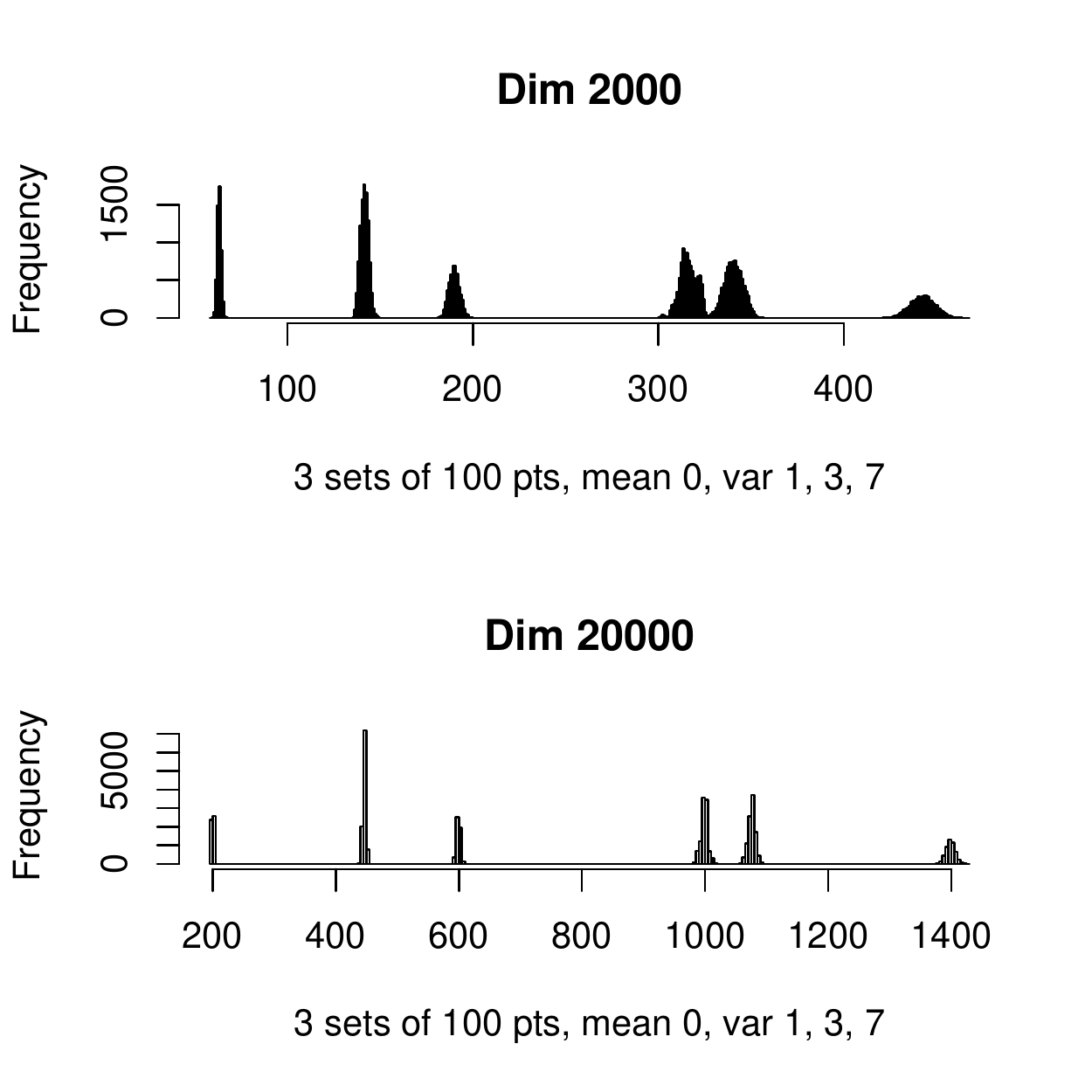}
\caption{An illustration of how ``symmetry'' or ``structure'' can become 
increasingly pronounced as dimensionality increases.  The abscissa shows 
distance values.  Displayed are two simulations, each with 3 sub-populations of 
Gaussian-distributed data,
in, respectively, ambient dimensions of 2000 and 20,000.  These simulations
correspond to the 3rd last, and 2nd last, rows of Table \ref{tabunifgauss}.}
\label{fig1}
\end{figure*}

Thus we see that in very high dimensions, and/or in very (spatially) 
sparse data clouds, 
there is a simplification of structure, 
which can be used to mitigate any ``curse of dimensionality''.  
Figure \ref{fig1}
shows how the distances within and between clusters become tighter with 
increase in 
dimensionality.  

\subsection{Gaussian Clusters in Very High Dimensions}
\label{sect33}

\subsubsection{Introduction}

We will distinguish between cluster characteristics as follows:

\begin{enumerate}
\item cluster size: number of points per cluster;
\item cluster location: here, mean, identical on every dimension;
\item cluster scale: here, standard deviation, identical on 
every dimension.
\end{enumerate}

These cluster characteristics are simple ones which serve to exemplify
how high dimensional clustering is quite different from analogous 
problems in low dimensions.  In the 
homogeneous clouds studied in Table \ref{tabunifgauss} it is seen that 
the isosceles (with small base) case disappeared early on, as dimensionality
increased greatly, to the advantage of the equilateral case of ultrametricity.
So the points become increasingly equilateral-related as dimensionality 
grows.  This is not the case when the data in clustered, as we will now
see.

\subsubsection{Clusters with Different Locations, Same Scale}

\begin{figure*}
\includegraphics[width=16cm]{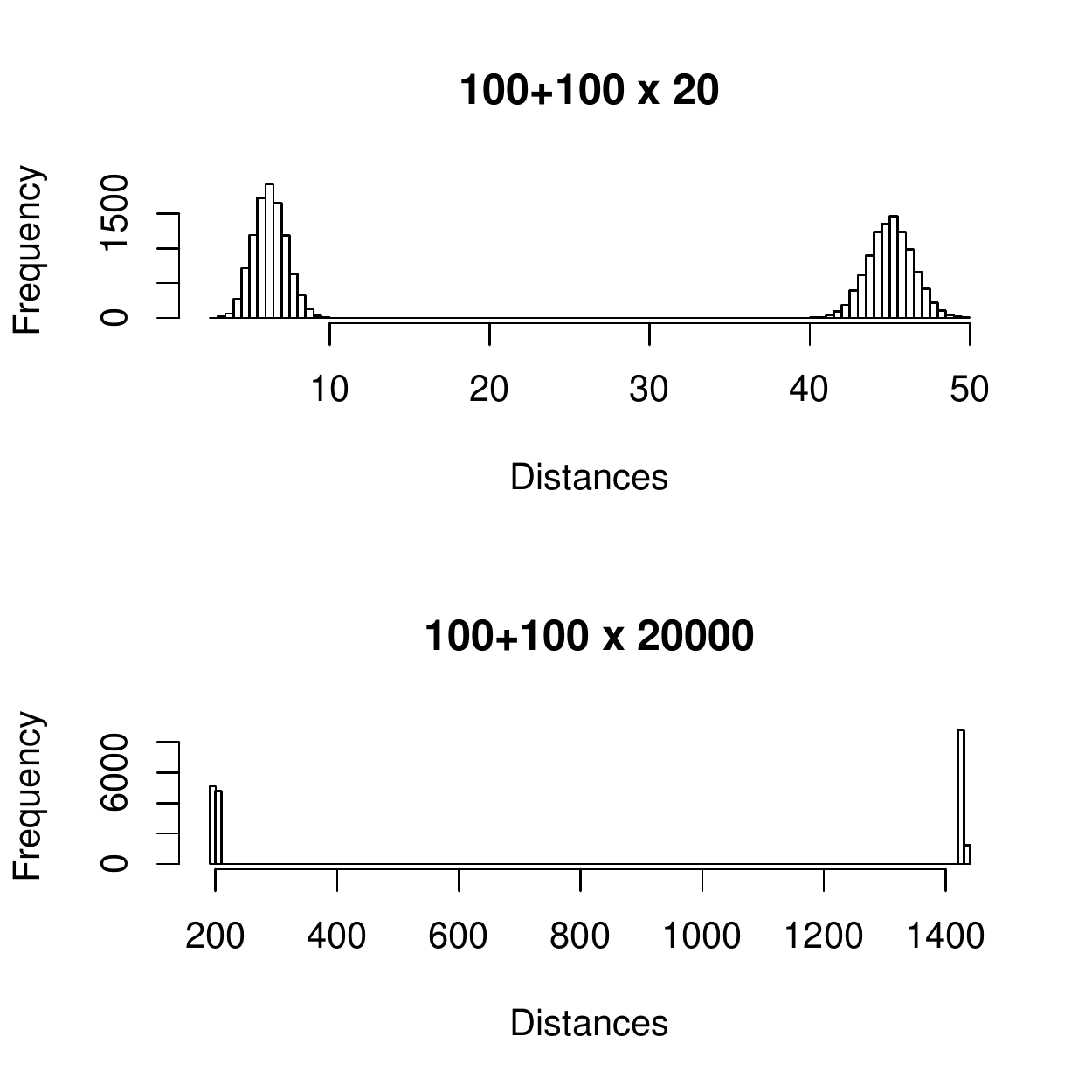}
\caption{A further 
illustration of how ``symmetry'' or ``structure'' can become 
increasingly pronounced as dimensionality increases, relating to the 
$200 \times 20$ and $200 \times 20,000$ (first of the succession of rows)
cases
of Table \ref{tabgauss}.  These are histograms of all interpoint
distances, based on two Gaussian clusters.  The first has mean 0 and standard
deviation 1 on all dimensions.  The second has mean 10 and standard deviation 
 1 on all dimensions.}  
\label{fig2}
\end{figure*}

Table \ref{tabgauss} is based on two clusters, and shows how isosceles
triangles increasingly dominate as dimensionality grows.  Figure \ref{fig2}
illustrates low and high dimensionality scenarios relating to Table
\ref{tabgauss}.  There is clear confirmation in this table as to how 
interrelationships in the cluster become more compact and, in a certain 
sense, more trivial, in high dimensions.  This does not obscure the 
fact that we indeed have hierarchial relationships becoming ever more 
pronounced as dimensionality, and hence relative sparsity, increase. 
These observations help us to see quite clearly just how hierachical 
relationships come about, as ambient dimensionality grows.  

\begin{table}
\begin{center}
\begin{tabular}{lllll} \hline
No. points &    Dimen. &  Isosc. &   Equil. &    UM   \\ \hline    
           &           &         &          &         \\
200        &    20     &    0.08 &    0  &    0.08 \\
200        &    200    &    0.19 &    0.04  &    0.23 \\
200        &    2000   &    0.42 &    0.20  &    0.62 \\
200        &    20000  &    0.74  &    0.22  &    0.96 \\ 
           &           &          &          &         \\
200        &    20000  &    0.7   &    0.28  &    0.98 \\ 
200        &    20000  &    0.77  &    0.21  &    0.98 \\ 
200        &    20000  &    0.76  &    0.21  &    0.98 \\ 
200        &    20000  &    0.75  &    0.24  &    0.99 \\ 
200        &    20000  &    0.73  &    0.25  &    0.98 \\ \hline
\end{tabular}
\end{center}
\caption{Results based on 300 sampled triangles from triplets of
points.  Two Gaussian clusters, each of 100 points,  were used in each case.
One point set was of mean 0, and the other of mean 10, on each dimension.
The standard deviations on each dimension were 1 in all cases.  Column 
headings are as in Table \ref{tabunifgauss}.  Five further results are given
for the 20,000-dimension case to show variability.}
\label{tabgauss}
\end{table}

\subsubsection{Clusters with Different Locations, Different Scales}

\begin{figure*}
\includegraphics[width=16cm]{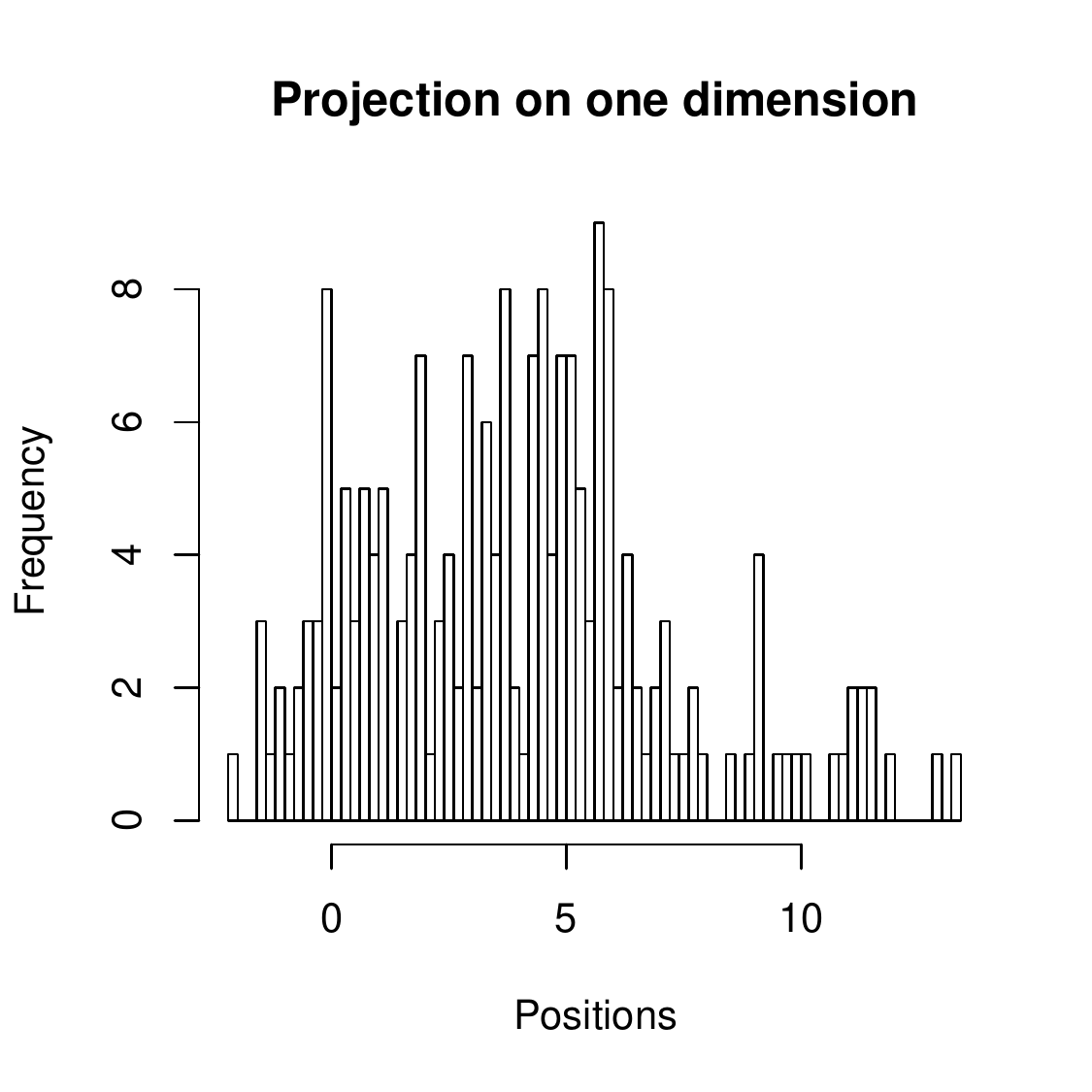}
\caption{A projection onto one dimension, to illustrate the less than
clearcut clustering problem addressed.  There are four Gaussians here,
each of 50 realizations, with means at 0, 3, 5 and 8, and with respective
standard deviations of 1, 2, 1, 3.}
\label{fig4}
\end{figure*}

A more demanding case study is now tried.  We generate 50 points per cluster
with the following characteristics: mean 0, standard deviation 1, on each 
dimension; mean 3, standard deviation 2, on each dimension; mean 5, standard
deviation 1, on each dimension; and mean 8, standard deviation 3, on each 
dimension.  Table \ref{tabgauss2} shows the results obtained.  Here 
we have not achieved quite the same level of ultrametricty, due to slower
growth in ultrametricity which is, in turn, due to the more murky, 
less dermarcated, but undoubtdely clustered, set of data.  Figure 
\ref{fig4} illustrates this: this histogram shows one dimension
(i.e., one coordinate, chosen arbitrarily), where 
we note that means of the Gaussians are at 0, 3, 5 and 8.  

\begin{table}
\begin{center}
\begin{tabular}{lllll} \hline
No. points &    Dimen. &  Isosc. &   Equil. &    UM   \\ \hline    
           &           &         &          &         \\
200        &    20     &    0.04 &    0.01  &    0.05 \\
200        &    200    &    0.11 &    0.05  &    0.16 \\
200        &    2000   &    0.28 &    0.06  &    0.34 \\
200        &    20000  &    0.5  &    0.08  &    0.58 \\
200        &    200000 &    0.55 &    0.11  &    0.66 \\ \hline
\end{tabular}
\end{center}
\caption{Results based on 300 sampled triangles from triplets of
points.  Four Gaussian clusters, each of 50 points,  were used in each case.
See text for details of properties of these clusters.} 
\label{tabgauss2}
\end{table}

\begin{figure*}
\includegraphics[width=16cm]{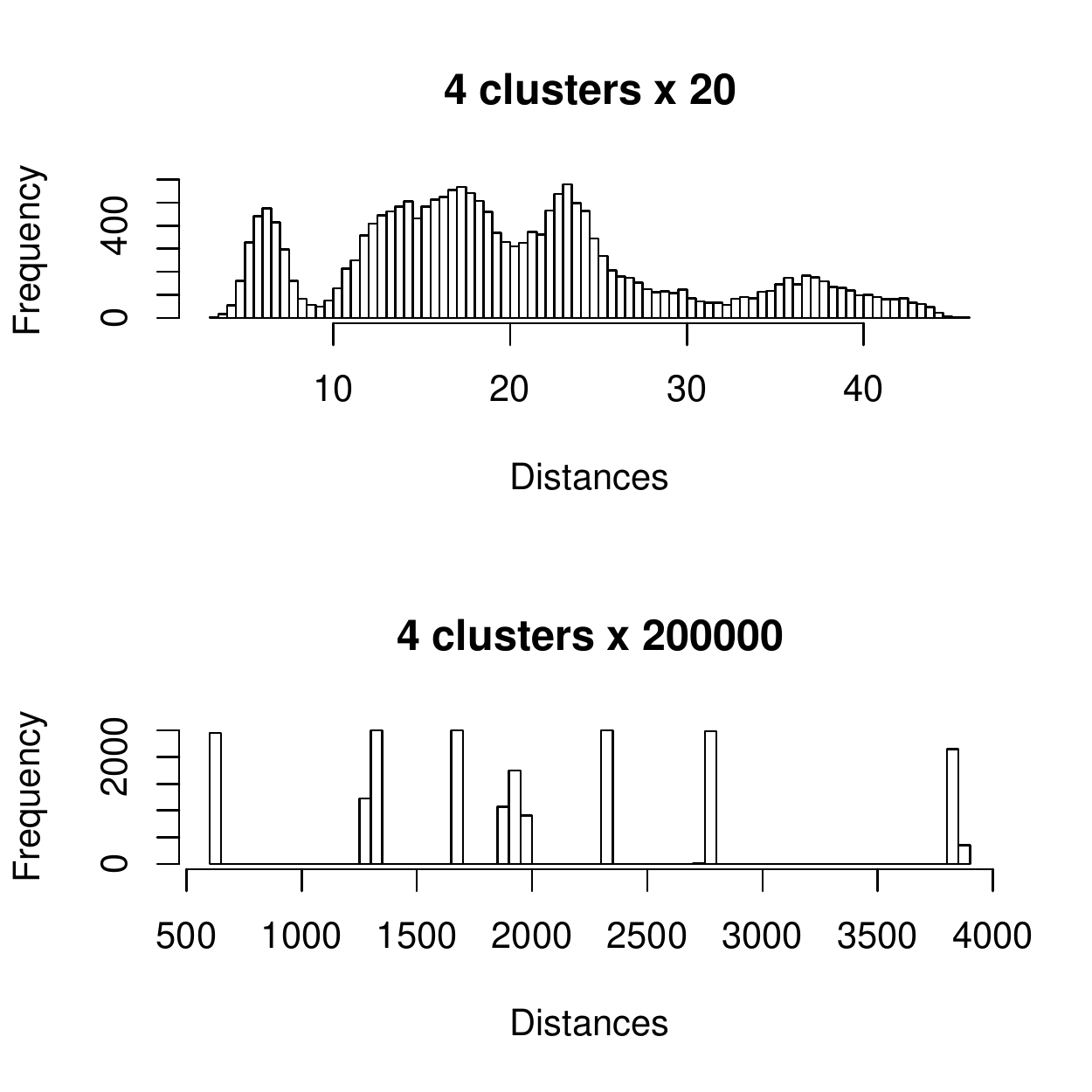}
\caption{Compaction of distances with rise in dimensionality: 4 clusters,
substantially overlapping are the basis for the histograms of all 
pairwise distances. Top: ambient dimensionality 20.  Bottom: ambient 
dimensionality 200,000.}
\label{fig3}
\end{figure*}

When we look closer at Table \ref{tabgauss2}, as shown in Figure 
\ref{fig3}, the compaction of distances is again very interesting.  
We verified the 7 peaks found in the lower histogram in Figure \ref{fig3},
and available but confusedly overlapping and ill-defined in the upper 
histogram of Figure \ref{fig3}.

What we find for the 7 peaks is as follows.  
Distances within the clusters correspond to: peaks 1,
2, 3 and (again) 1.  That two clusters are associated with one peak is
clear from the fact that two of our clusters are of identical scale.  

We can examine inter-cluster distances and we found these to be associated
with peaks: 2, 3, 4, 5, 6, 7.  Given 4 clusters, we could well have up to 
6 possible additional peaks.  

Were we to be in a far higher dimensional ambient space then we could expect
even the inter-cluster distances to become equi-distant.  

%


\subsubsection{Identifiability of High Dimensional Gaussian Clouds}

From these case studies, it is clear that increased dimensionality 
sharpens and distinguishes the clusters.  
If we can embed data -- any data -- in a far
higher ambient dimensionality, without destroying the interpretable 
relationships in the data, then we can so much more easily read off the 
clusters.  

To read off clusters, including memberships and properties, our findings
can be summarized as follows.  

For cluster size (i.e., numbers of points per cluster), sampling alone 
can be used, and we do not pursue this here.  

For cluster scale (i.e., standard deviation, assumed the same on each 
dimension), we associate each cluster, or a pair of clusters, with each
peak.    The total number of peaks gives an upper bound on the number of
clusters.  (For $k$ clusters, we have $ \leq k + k \cdot (k-1) /2$ 
peaks.) 

Using cluster scale also permits use of the following cluster model: 
suppose that all clusters are defined to have intra-cluster distance 
that is less than inter-cluster distance.  Then it follows that the 
peaks of lower distance correspond to the clusters (as opposed to pairs
of clusters).  

An example of this is as follows.  
In Figure \ref{fig3}, lower panel, we read from left to 
right, applying the following algorithm: select the first $k$ peaks as 
clusters, and ask: are there sufficient peaks to represent all 
inter-cluster pairs?  If we choose $k = 3$, there remain 4 peaks, which is
too many to account for the inter-cluster pairs (i.e., $ 3 \cdot (3-1)/2)$).  
So we see that Figure \ref{fig3} is incompatible with 
$k = 3$ or the presence of just 3 clusters.  

Consequently we move to $k = 4$, and see that Figure \ref{fig3} is 
consistent with this.  

A further identifiability assumption is reasonable albeit not required:
that all smallest peaks be associated with intra-cluster distances.  
This need not be so, since we could well have a dense cluster superimposed on
a less dense one.  However it is a reasonable parsimony assumption.  Supported
by this assumption, Figure \ref{fig3} points to a minimum of 4 clusters in 
the data, with up to 4 peaks (read off from left to right, i.e., in increasing
order of distance) corresponding to these clusters.  


\begin{figure*}
\includegraphics[width=16cm]{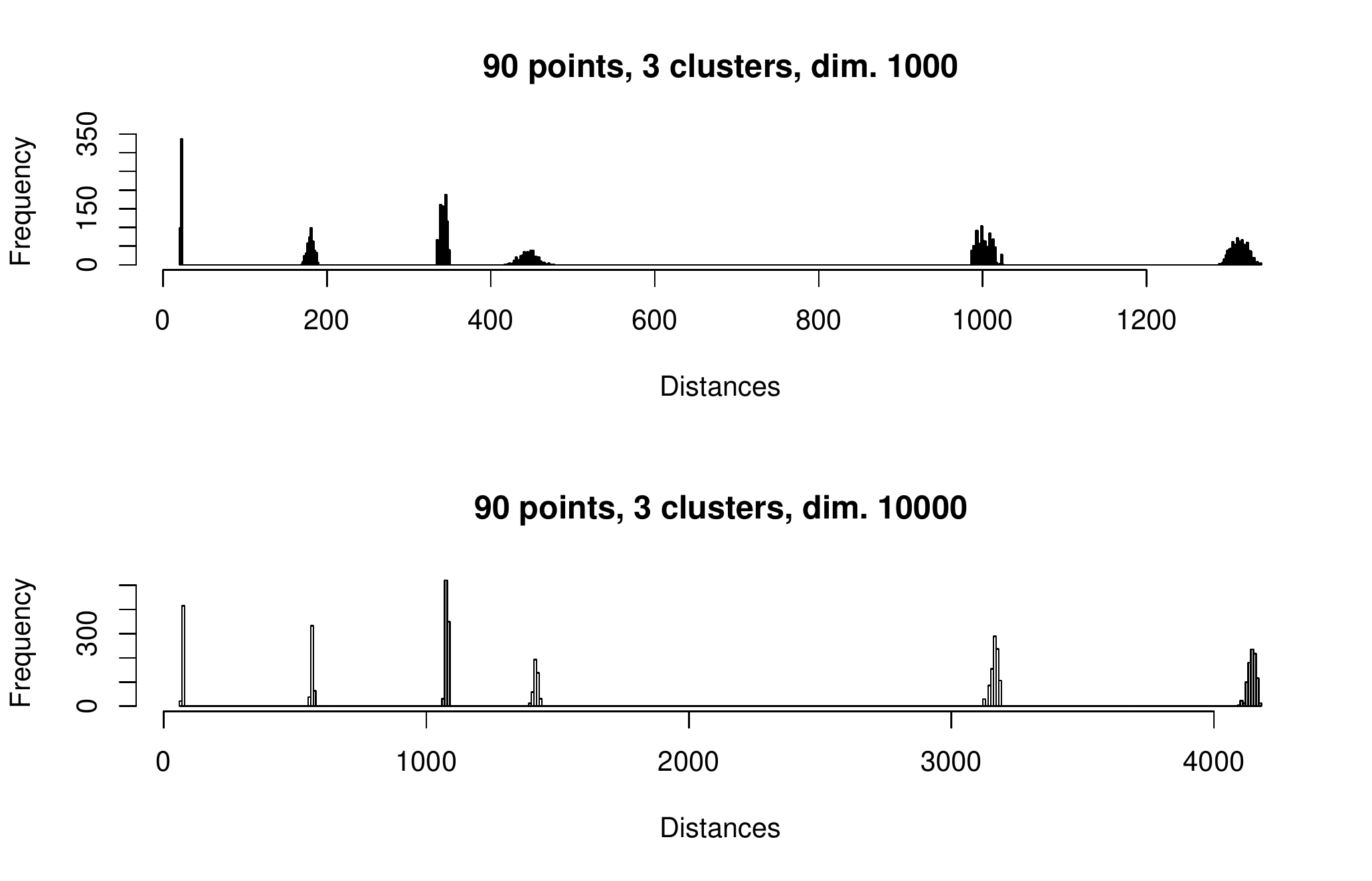}
\caption{Histogram of interpoint distances.  
Three homogeneous clusters, each of 30 points, in spaces
of dimensions 1000 and 10000.}
\label{fig30}
\end{figure*}

\begin{figure*}
\includegraphics[width=16cm]{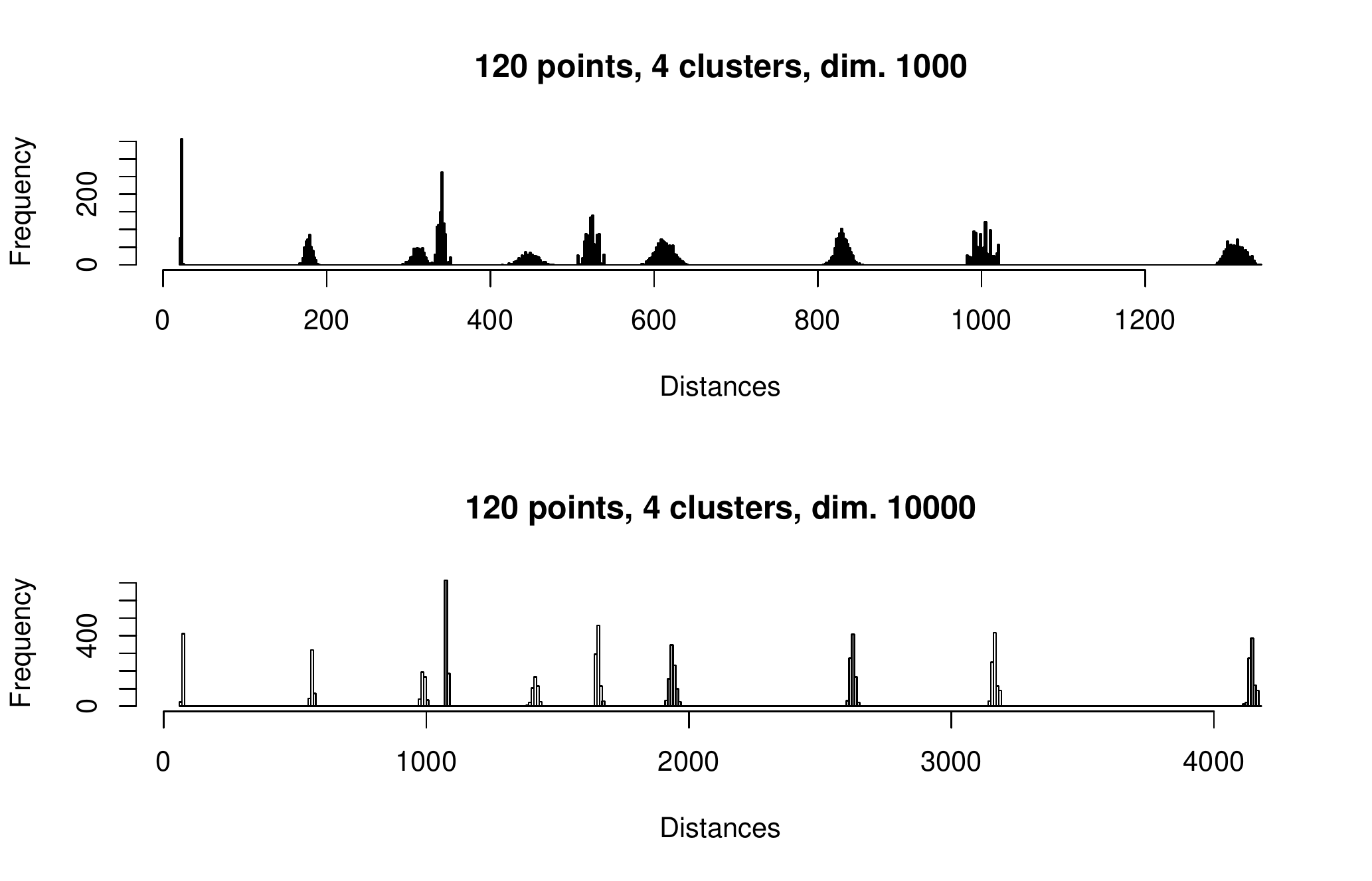}
\caption{Histogram of interpoint distances.  
Four homogeneous clusters, each of 30 points, in spaces
of dimensions 1000 and 10000.}
\label{fig32}
\end{figure*}

\begin{figure*}
\includegraphics[width=16cm]{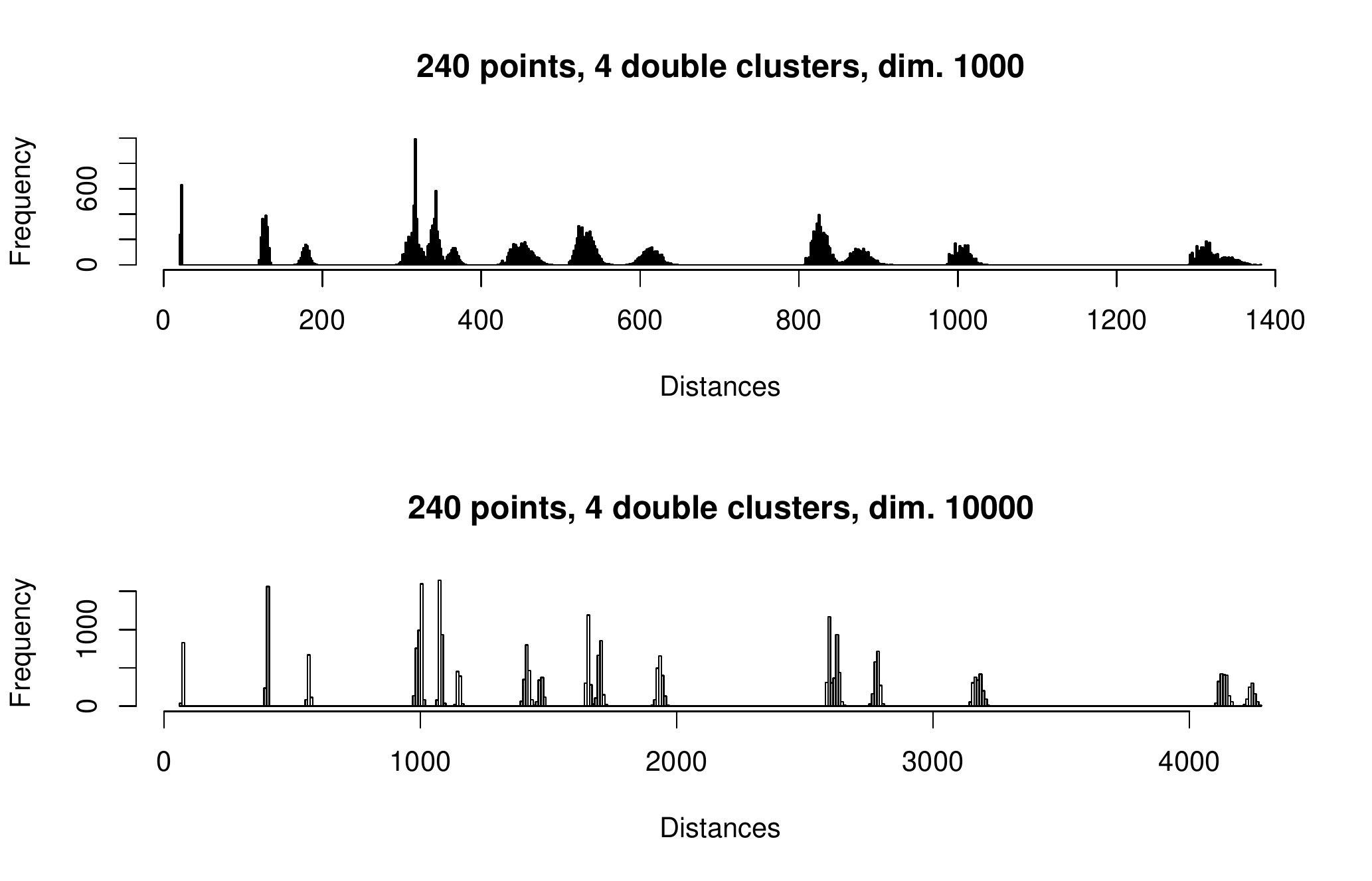}
\caption{Histogram of interpoint distances.  
Four heterogeneous clusters, each of 60 points (comprising two subgroups of
30 points each), in spaces
of dimensions 1000 and 10000.}
\label{fig34}
\end{figure*}

Figure \ref{fig30} shows peaks, sharpening with rise in ambient dimensionality,
for three clusters, distributed as Gaussians with respectively means and 
standard deviations on all dimensions: (10, 0.5); (0, 4); (40, 10).  We see
the peaks corresponding to the increasingly similar (and tending towards identical)
intra-cluster distances; and then the peaks associated with the 
3 inter-cluster distance sets: 6 peaks in total.  

Figure \ref{fig32} again shows peaks for four clusters, with the same 
characteristics  as for Figure \ref{fig30}, but with an additional cluster 
of mean and standard deviation on all coordinates: (25, 7).  Here we find, 
as expected, 4 intra-cluster distance peaks, and 6 inter-cluster distance peaks:
10 peaks in total.

The success of cluster identification is clearly dependent on distinguishable
intra-cluster properties, which are also distinguishable from the inter-cluster
properties.  

Figure \ref{fig34} shows a more tricky case.  For the first cluster, the 
(equally-sized component) distributions had means and standard deviations on 
all coordinates of: (10, 0.5) and (0, 0.5).  For the second cluster, we used:
(0, 4) and (10, 4).  For the third cluster we used (40, 10) and (0, 10).
Finally for the fourth cluster we used the same for all cluster member points:
(25, 7).  Here, therefore the maximum number of peaks is: for the intra-cluster
distances, 2 peaks for the first, second and third clusters, and 1 for the 
fourth cluster, hence 7.  For the inter-cluster distances, assuming the 
cluster components are close enough, 6, but if they are not, then, we have 
peaks between essentially 7 clusters, i.e.\ 21 peaks.  Hence in total we 
could have up to 28 peaks.  For the histogram sampling resolution used, 
we read off, visually, 17 peaks.  

Our approach is to have an upper bound on the number of peaks found in the 
distance histogram.  
Now we turn to real (or realistic) data and see how this work helps us to 
address the cluster identifiability problem.

\section{Applications}
\label{appsect4}

\subsection{Application to High Frequency Data Analysis}
\label{hfda}

In this section we establish proof of concept for application of 
the foregoing work to analysis of very high frequency 
univariate time series signals.  

Consider each of the cases considered in section \ref{sect33},
expressed there as $n \times m$ arrays, as instead representing $n$ segments,
each of (contiguous) length $m$, of a time series or one-dimensional signal.
Assuming our aim is to cluster these segments on the basis of their 
properties, then it is reasonable to require that they be non-overlapping.  
The $n$ segments could come from anywhere, in any order, in the time series.
So for the case of an $n \times m$ array considered previously, then 
implies a time series of length at least $n m$.  The most immediate way 
to construct the time series is to raster scan the $n \times m$ array, 
although alternatives come readily to mind.  

The methodology discussed in section \ref{sect33} then is seen to be also
a time series segmentation approach, facilitating the characterizing of 
the segments used.  

\begin{figure*}
\includegraphics[width=16cm]{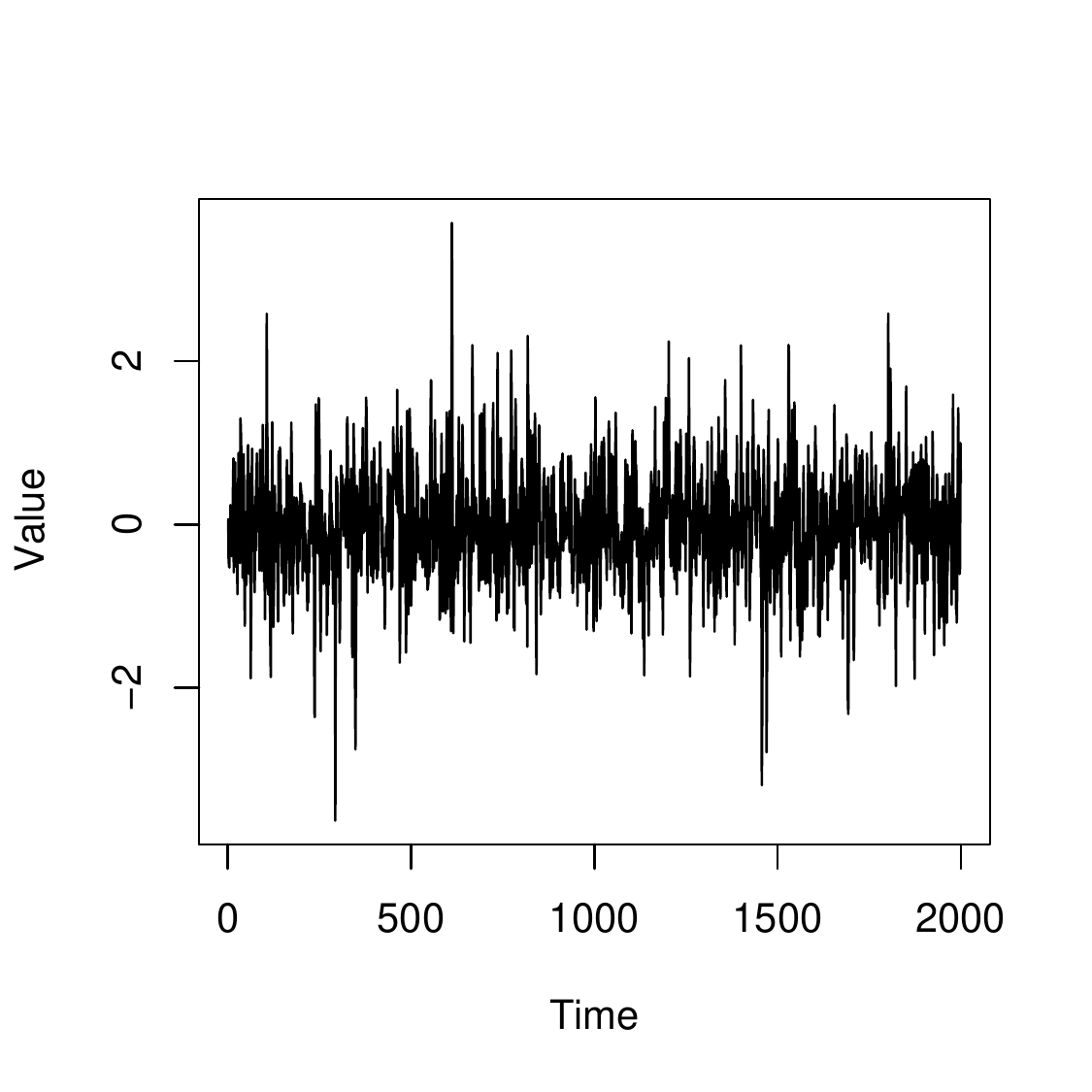}
\caption{Sample (using first 2000 values) of a time series segment, based
on the first ARIMA set of parameters. (Order 2 
AR parameters: $0.8897, -0.4858$,
MA parameters: $-0.2279, 0.2488$.)}
\label{fig6}
\end{figure*}

\begin{figure*}
\includegraphics[width=16cm]{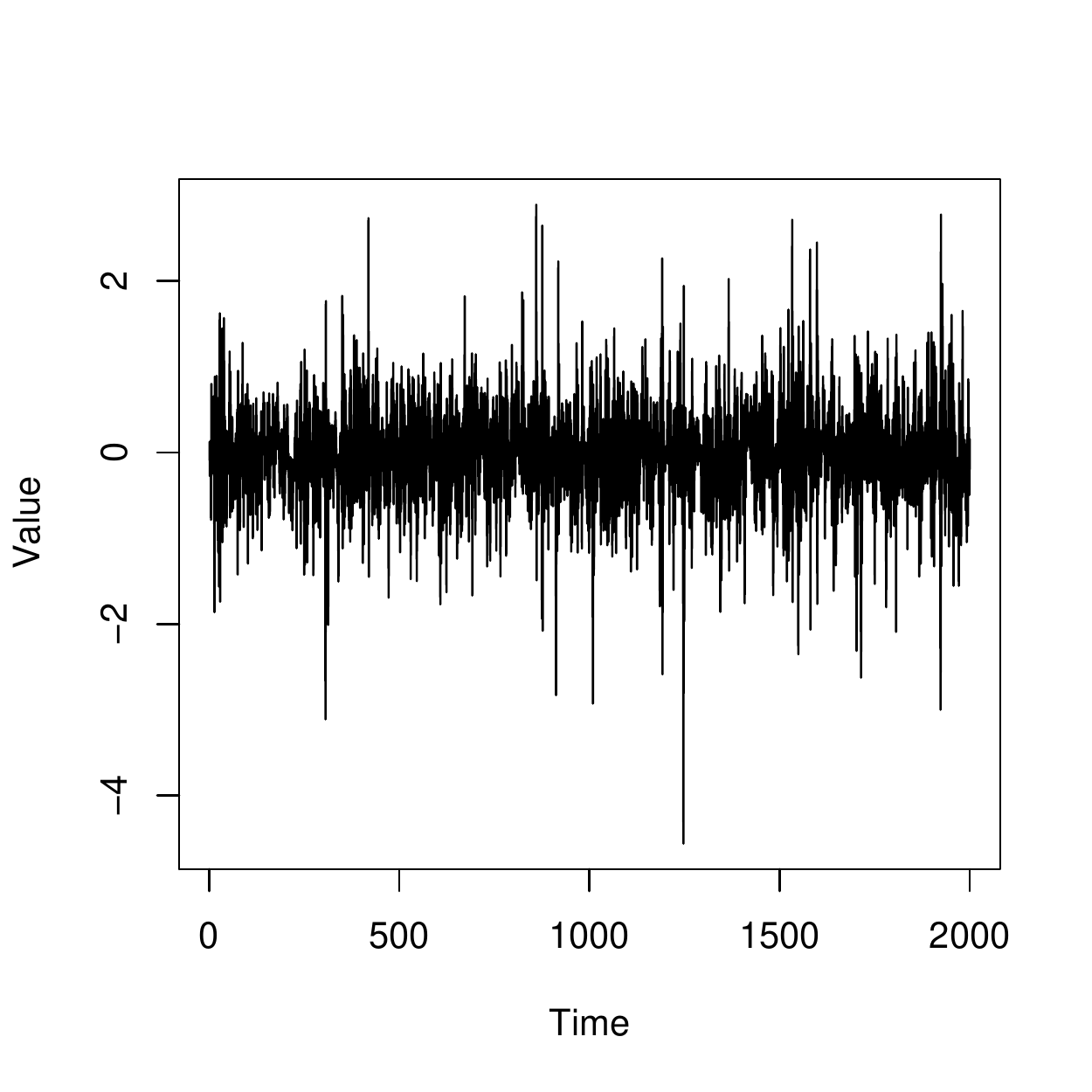}
\caption{Sample (using first 2000 values) of a time series segment, based
on the second ARIMA set of parameters. (Order 2
AR parameters: $0.2897, -0.1858$,
MA parameters: $-0.7279, 0.7488$.)}
\label{fig7}
\end{figure*}

\begin{figure*}
\includegraphics[width=16cm]{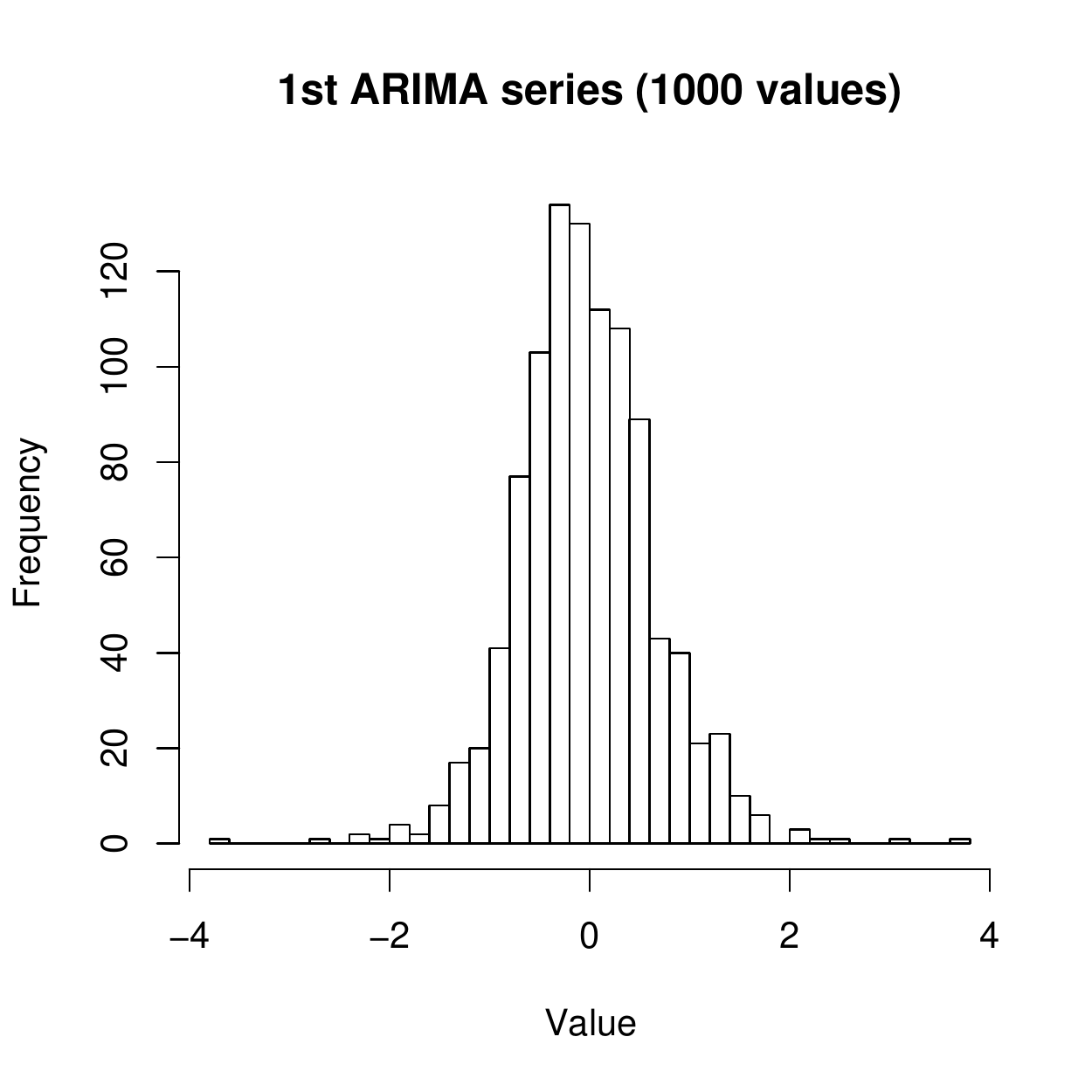}
\caption{Histogram of sample (using first 2000 values) of 
time series segment shown in Figure \ref{fig6}.}
\label{fig8}
\end{figure*}

\begin{figure*}
\includegraphics[width=16cm]{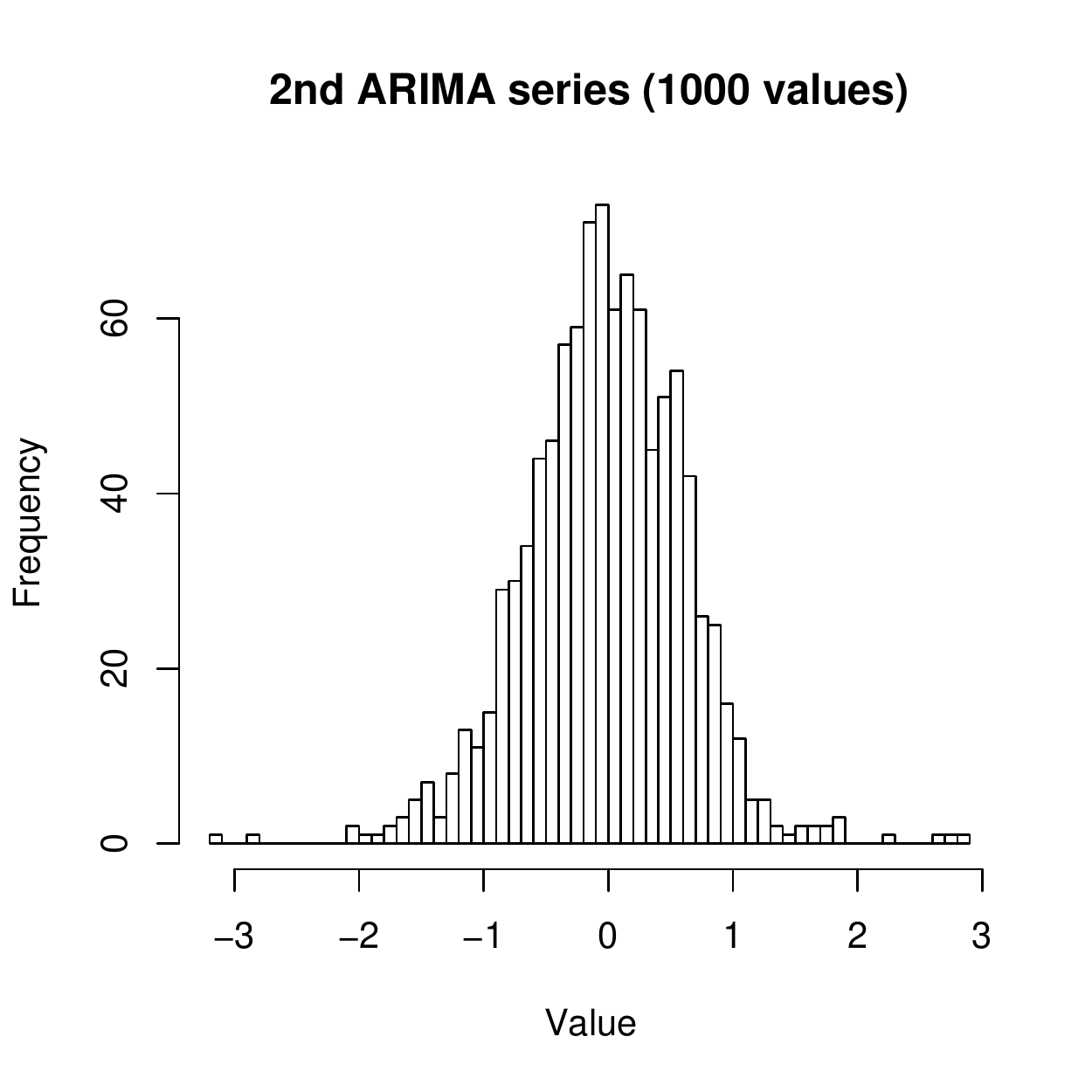}
\caption{Histogram of sample (using first 2000 values) of 
time series segment shown in Figure \ref{fig7}.}
\label{fig9}
\end{figure*}

To explore this further we consider a time series consisting of two 
ARIMA (autoregressive integrated moving average) models, with parameters:
order, autoregression coefficients, moving average coefficients, and a 
``mildly longtailed'' set of innovations based on the Student t distribution 
with 5 degrees of freedom.  
Figures \ref{fig6} and \ref{fig7} show samples of these time series 
segments.  Figures \ref{fig8} and \ref{fig9} show histograms of these 
samples. 

Table \ref{tabfindat} shows typical results obtained in regard to 
ultrametricity.  The dimensionality can be considered as the embedding
dimension.  Here, although ultrametricity increases, and the 
equilateral configuration seems to be increasing but with decrease of the 
isosceles with small base configuration, we do not consider it of 
practical relevance to test with even higher ambient dimensionalities.  
It is clear from the data, especially Figures \ref{fig8} and \ref{fig9},
that the two signal models are very close in their properties.  

\begin{figure*}
\includegraphics[width=16cm]{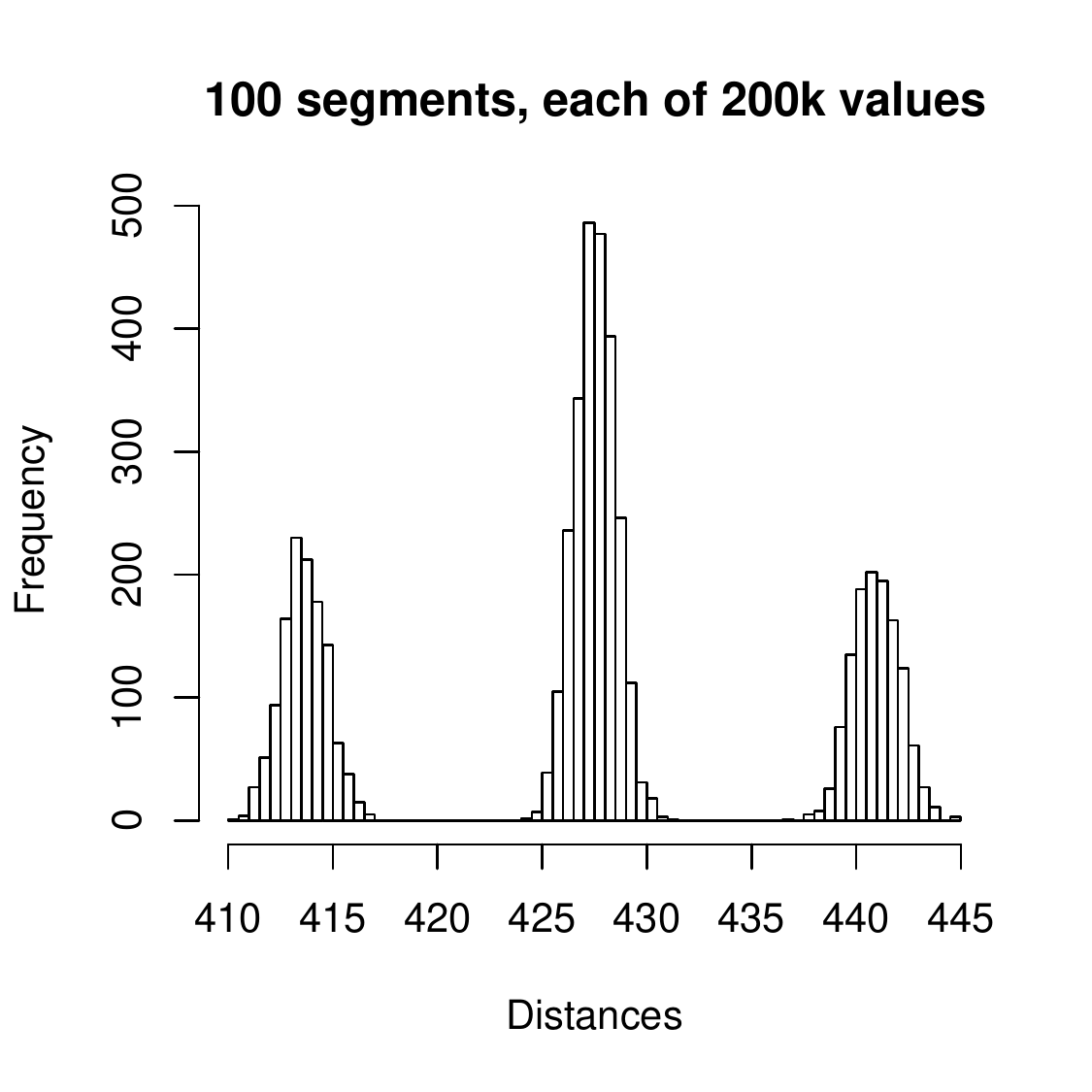}
\caption{Histogram of distances from 100 time series segments, 
using 50 segments each from the two ARIMA models, and using an 
embedding dimensionality of 200,000.}
\label{fig10}
\end{figure*}

Examining the histograms of all inter-pair time series segments, both 
intra and inter cluster, we find the clearly distinguished peaks shown 
in Figure \ref{fig10}.  As before, we use Euclidean distance between 
time series segments or vectors.  (We note that normalization or other 
transformation is not relevant here.  In fact we want to 
distinguish between inter and intra cluster cases.  Furthermore the 
unweighted Euclidean distance is consistent with our use of angles to 
quantify triangle invariants, and hence respect for ultrametricity properties.)

\begin{table}
\begin{center}
\begin{tabular}{lllll} \hline
No. time series &    Dimen. &  Isosc. &   Equil. &    UM   \\ \hline    
           &           &         &          &         \\
100        &    2000   &    0.17 &    0.32  &    0.49 \\
100        &    20000  &    0.15  &    0.5  &    0.65 \\
100        &    200000 &    0.03 &    0.57  &    0.60 \\ \hline
\end{tabular}
\end{center}
\caption{Results based on 300 sampled triangles from triplets of
points.  Two sets of the ARIMA models are used, each of 50 realizations.}
\label{tabfindat}
\end{table}

We find clearly distinguishable peaks in Figure \ref{fig10}.  The lower and 
the higher peaks belong to the two ARIMA components.  The central peak 
belongs to the inter-cluster distances.  

We have shown that our methodology can be of use for time series segmentation
and for model identifiability.  
Given the use of a scalar product space as the essential springboard of 
all aspects of this work, it would appear that generalization of this work 
to multivariate time series analysis is straightforward.  What remains 
important, however, is the availability of very large embedding 
dimensionalities, i.e.\ very high frequency data streams.  

\subsection{Application in Practice: Segmenting a Financial Signal}

\begin{figure*}
\includegraphics[width=16cm]{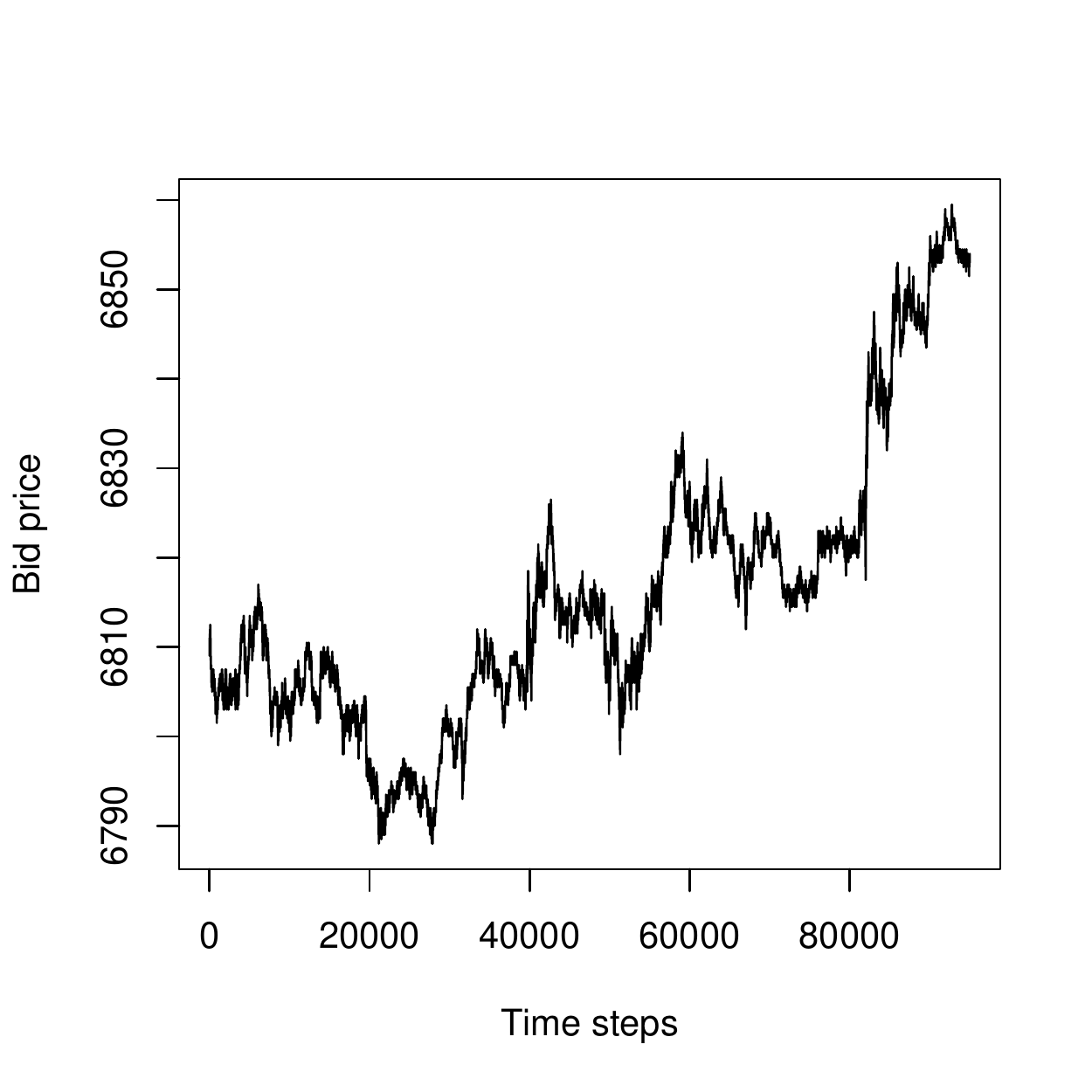}
\caption{The signal used: a commodity future, with millisecond time sampling.}
\label{fig100}
\end{figure*}

We use financial futures, circa March 2007, denominated in euros from the DAX
exchange.  Our  data stream is at the millisecond rate, and comprises about 382,860
records.  Each record includes: 5 bid and 5 asking prices, together with
bid and asking sizes in all cases, and action.  We extracted one symbol 
(commodity) with 95,011 single bid values, on which we now report results.  
See Figure \ref{fig100}.

Embeddings were defined as follows.  

\begin{itemize}
\item Windows of 100 successive values, starting at time steps: 
1, 1000, 2000, 3000, 4000, $\dots$, 94000.
\item Windows of 1000 successive values, starting at time steps: 
1, 1000, 2000, 3000, 4000, $\dots$, 94000.
\item Windows of 10000 successive values, starting at time steps: 
1, 1000, 2000, 3000, 4000, $\dots$, 85000.
\end{itemize}

\begin{figure*}
\includegraphics[width=16cm]{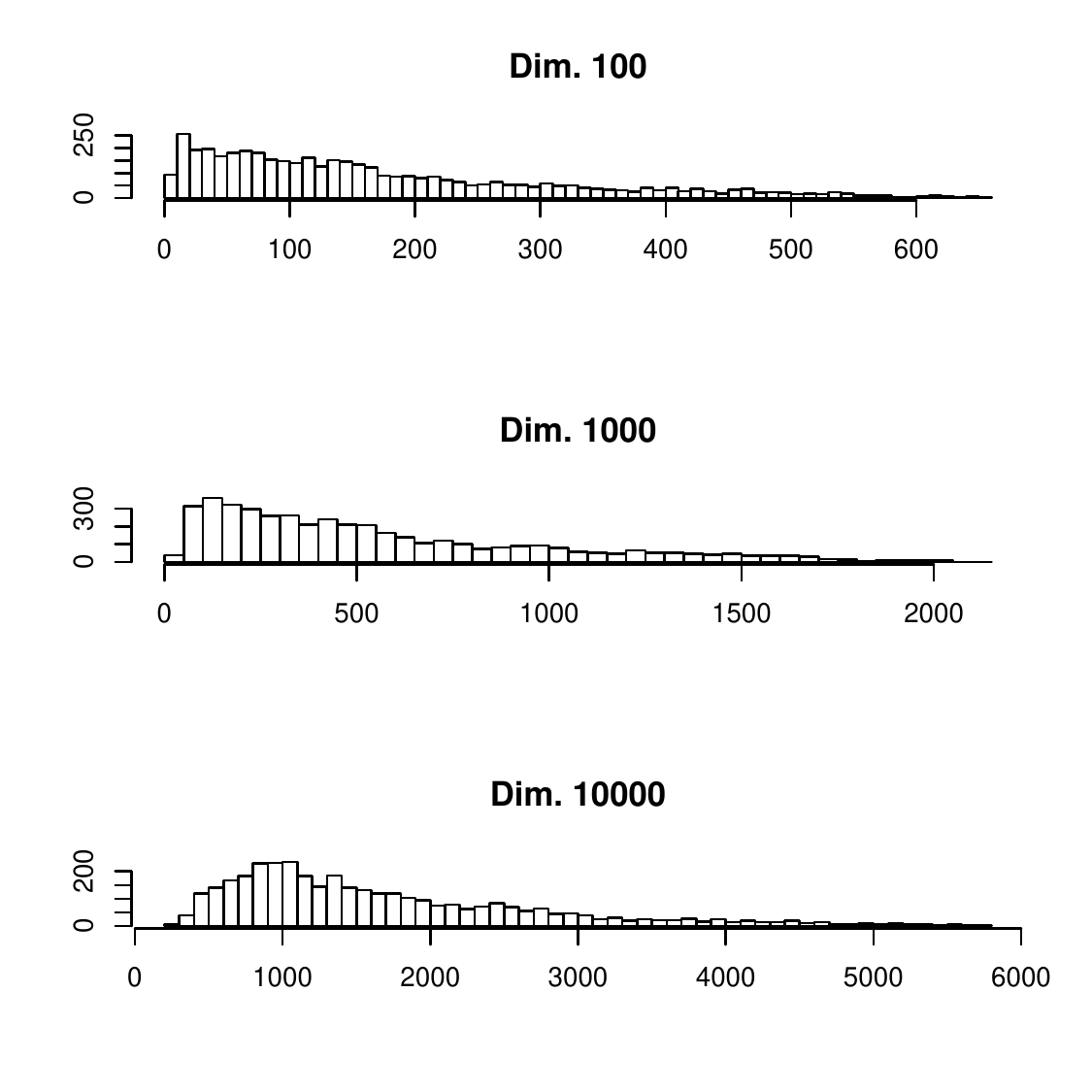}
\caption{Histograms of pairwise distances between 
embeddings in dimensionalities 100, 1000, 10000. 
Respectively the numbers of embeddings are: 95, 95 and 86.}
\label{fig110}
\end{figure*}

The histograms of distances between these windows, or embeddings, in respectively 
spaces of dimension 100, 1000 and 10000, are shown in Figure \ref{fig110}.

Note how the 
10000-length window case results in points that are strongly overlapping.
In fact, we can say that 90\% of the values in each window are overlapping 
with 
the next window.  Notwithstanding this major overlapping in regard to clusters
involved in the pairwise distances, if we can still find clusters in the 
data then we have a very versatile way of tackling the clustering objective.
Because of the greater cluster concentration that we expect (from 
discussion in earlier sections of this article) from a greater embedding
dimension, we use the 86 points in 10000-dimensional space, notwithstanding 
the fact that these points are from overlapping clusters.  

We make the following supposition based on Figure \ref{fig100}: the 
clusters will consist of successive values, and hence will be justifiably 
termed segments.  

To validate our approach we will pursue three separate attacks on the
same problem of time series segmentation.  Firstly, from the distances
histogram in Figure \ref{fig110}, bottom, we will carry out Gaussian 
mixture modeling followed by use of the Bayesian information criterion (BIC,
Schwarz, 1978) 
as an approximate Bayes factor, to determine the best number of 
clusters (effectively, histogram peaks).  Secondly we will use an
adjacency-respecting hierarchical clustering algorithm on the 
full-dimensional (viz., 10000) data.  Thirdly, we will use a 
reduced dimensionality mapping, principal coordinates analysis, 
using the inter-point distances.
Our assumptions in regard to what clusters are present in the data are
minimal.  Furthermore our validation of segments is based on the 
three different ways that we have of tackling the one segmentation problem.

\begin{figure*}
\includegraphics[width=16cm]{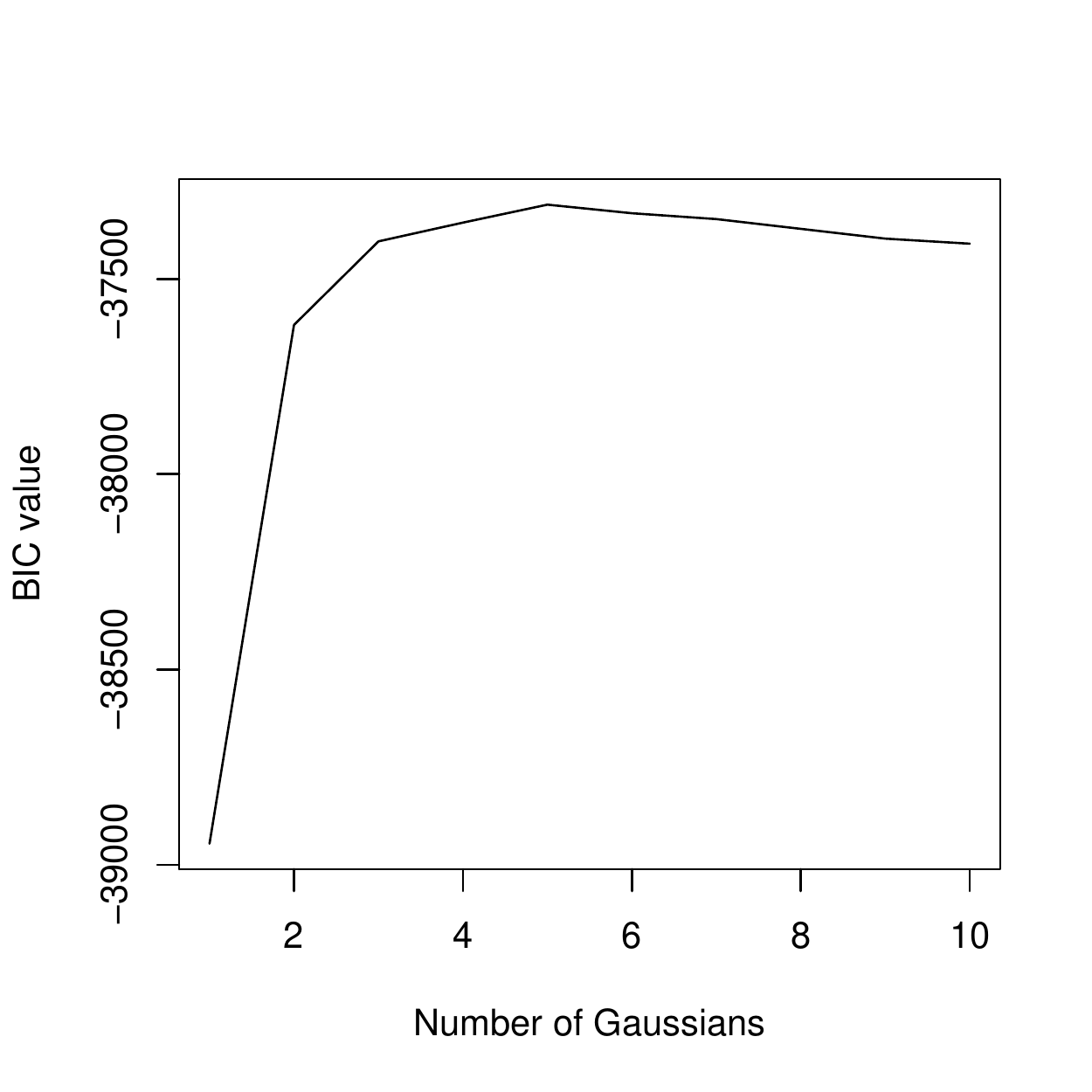}
\caption{BIC (Bayesian information criterion) values for the succession of 
results.  The 5-cluster solution has the highest value for BIC and is therefore
the best Gaussian mixture fit.}
\label{fig120}
\end{figure*}

We fit a Gaussian mixture model 
to the data shown in the bottom histogram of Figure \ref{fig110}.  
To derive the appropriate number of histogram peaks we fit Gaussians and use 
the Bayesian information criterion (BIC) as an approximate Bayes factor for 
model selection (Kass and Raftery, 1995; Murtagh and Starck, 2003).  
Figure \ref{fig120} shows the 
succession of outcomes, and indicates as best a 5-Gaussian fit.
For this result, we find  the means of the Gaussians 
to be as follows:  517, 885, 1374, 2273 and 3908.  The corresponding 
standard deviations are: 84, 133, 212, 410 and 663.  The respective 
cardinalities
of the 5 histogram peaks are: 358, 1010, 1026, 911 and 350.  Note that 
this relates
so far only to the histogram of pairwise distances.  We now want to determine
the corresponding clusters in the input data.  

While we have the segmentation of the distance histogram, we need the
segmentation of the original financial signal.  If we had 2 clusters in the 
original financial signal, then we could expect up to 3 peaks in the
distances histogram (viz., 2 intra-cluster peaks, and 1 inter-cluster peak).
If we had 3 clusters in the original financial signal, then we could 
expect up to 6 peaks in the distances histogram (viz., 3 intra-cluster
peaks, and 3 inter-cluster peaks).  This information is consistent 
with asserting that the evidence from Figure \ref{fig120} points to 
two of these histogram peaks being approximately co-located (alternatively:
the distances are approximately the same).  We conclude that 3 clusters
in the original financial signal is the most consistent number of clusters.
We will now determine these.  

\begin{figure*}
\includegraphics[width=16cm]{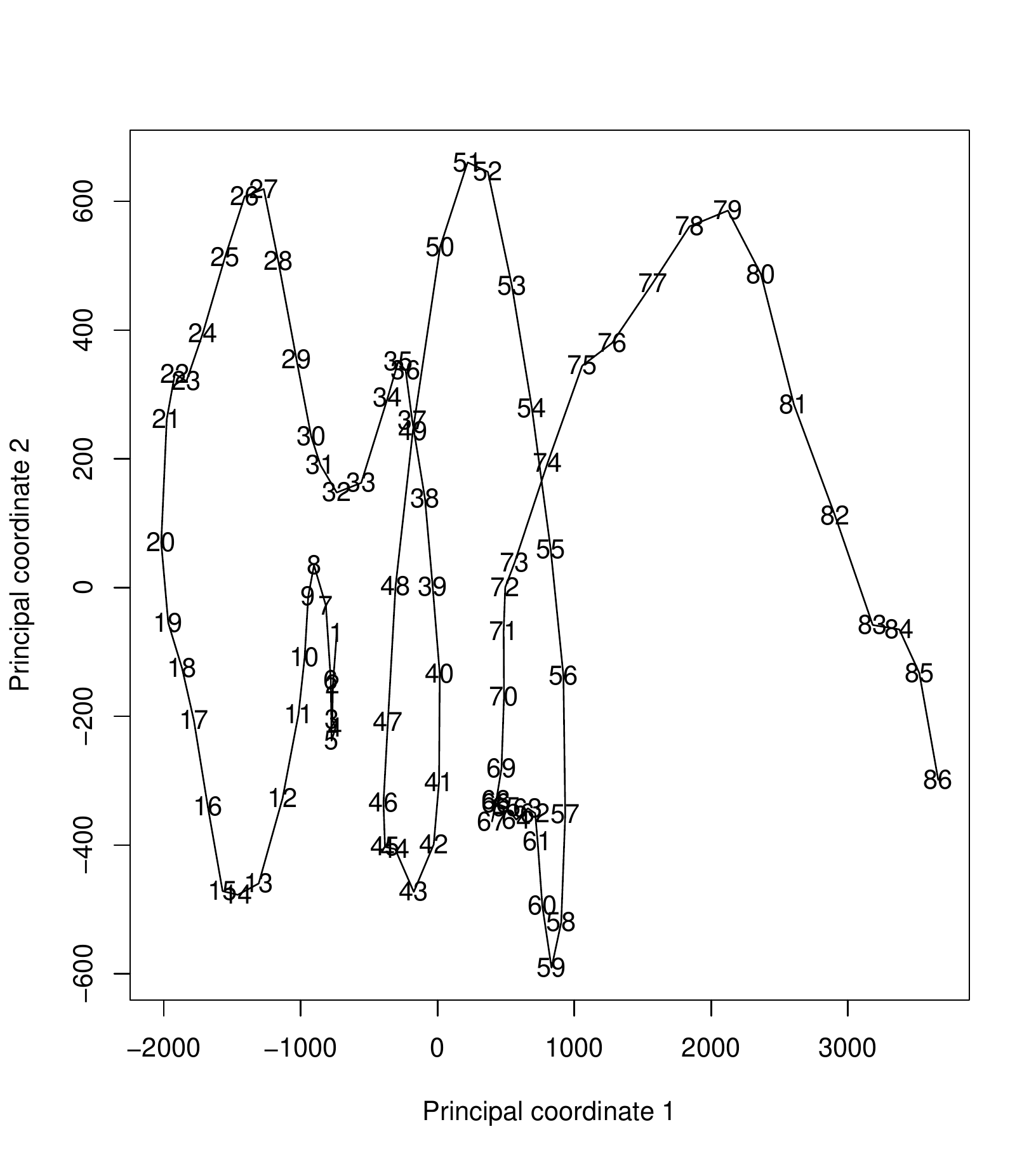}
\caption{An interesting representation -- a type of ``return map'' -- found
using a principal coordinates analysis of the 86 successive 
 10000-dimensional points.
Again a demonstration that very high dimensional structures can be of 
very simple structure.  The planar projection seen here represents most 
of the information content of the data: the first axis accounts for 
88.4\% of the variance, while the second accounts for 5.8\%.}
\label{fig180}
\end{figure*}

One possibility is to use principal coordinates analysis 
(Torgerson's, Gower's metric multidimensional scaling) of the pairwise
distances.  In fact, a
2-dimensional mapping furnishes a very similar pairwise
distance histogram to that seen using
the full, 10000, dimensionality.  The first axis in Figure \ref{fig180} 
accounts for 88.4\% of the variance, and the second for 5.8\%.
Note therefore how the scales of the planar representation in Figure 
\ref{fig180} point to it being very linear.  

Benz\'ecri (1979, Vol.\ II, chapter 7,
section 3.1) discusses the Guttman effect, or Guttman scale, where 
factors that are not mutually correlated, are nonetheless
functionally related.  When there is a ``fundamentally unidimensional 
underlying phenomenon'' (there are multiple such cases here) 
factors are functions of Legendre polynomials. 
We can view Figure \ref{fig180} as consisting of multiple horseshoe shapes.
A simple explanation for such shapes is in terms of the 
constraints imposed by lots of equal distances when the data vectors are
ordered linearly: see Murtagh (2005, pp.\ 46-47).

Another view of how embedded (hence clustered) data are capable of 
being well mapped into a unidimensional curve is Critchley and Heiser
(1988).  Critchley and Heiser show one approach to mapping an ultrametric into 
a linearly or totally ordered metric.  We have asserted and then established
how hierarchy in some form is relevant for high dimensional data spaces; and
then we find a very linear projection in Figure \ref{fig180}.  As a consequence
we note that the Critchley and Heiser result is especially relevant for 
high dimensional data analysis.  

\begin{figure*}
\includegraphics[width=20cm,angle=270]{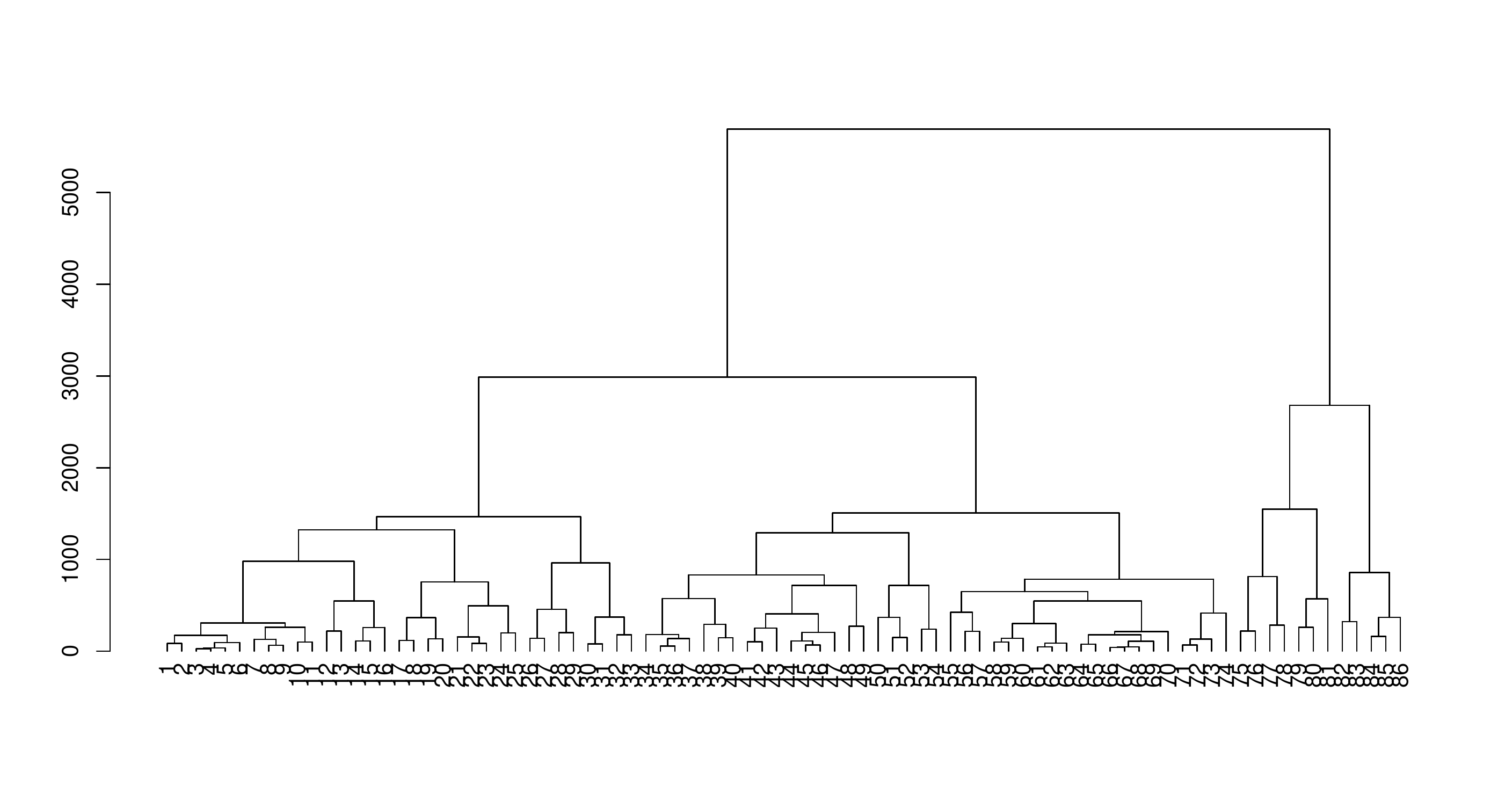}
\caption{Hierarchical clustering of the 86 points.  
Sequence is respected.  The agglomerative criterion is the 
contiguity-constrained complete link method.  See Murtagh (1985) for 
details including proof that there can be no inversion in this dendrogram.}
\label{fig140}
\end{figure*}

Knowing that 3 clusters in the original signal are wanted, 
we will use an adjacency-constrained agglomerative
hierarchical clustering algorithm to find them:
see Figure \ref{fig140}.  The contiguity-constrained 
complete link criterion 
is our only choice here if we are to be sure that no inversions 
can come about in the hierarchy, as explained in Murtagh (1985).  
As input, we use the coordinates in Figure \ref{fig180}.
The 2-dimensional Figure \ref{fig180} representation relates
to over 94\% of the variance.  The most complete basis was of 
dimensionality 85.  We checked the results of the 85-dimensionality 
embedding which, as noted below, gave very similar results.  

\begin{figure*}
\includegraphics[width=16cm]{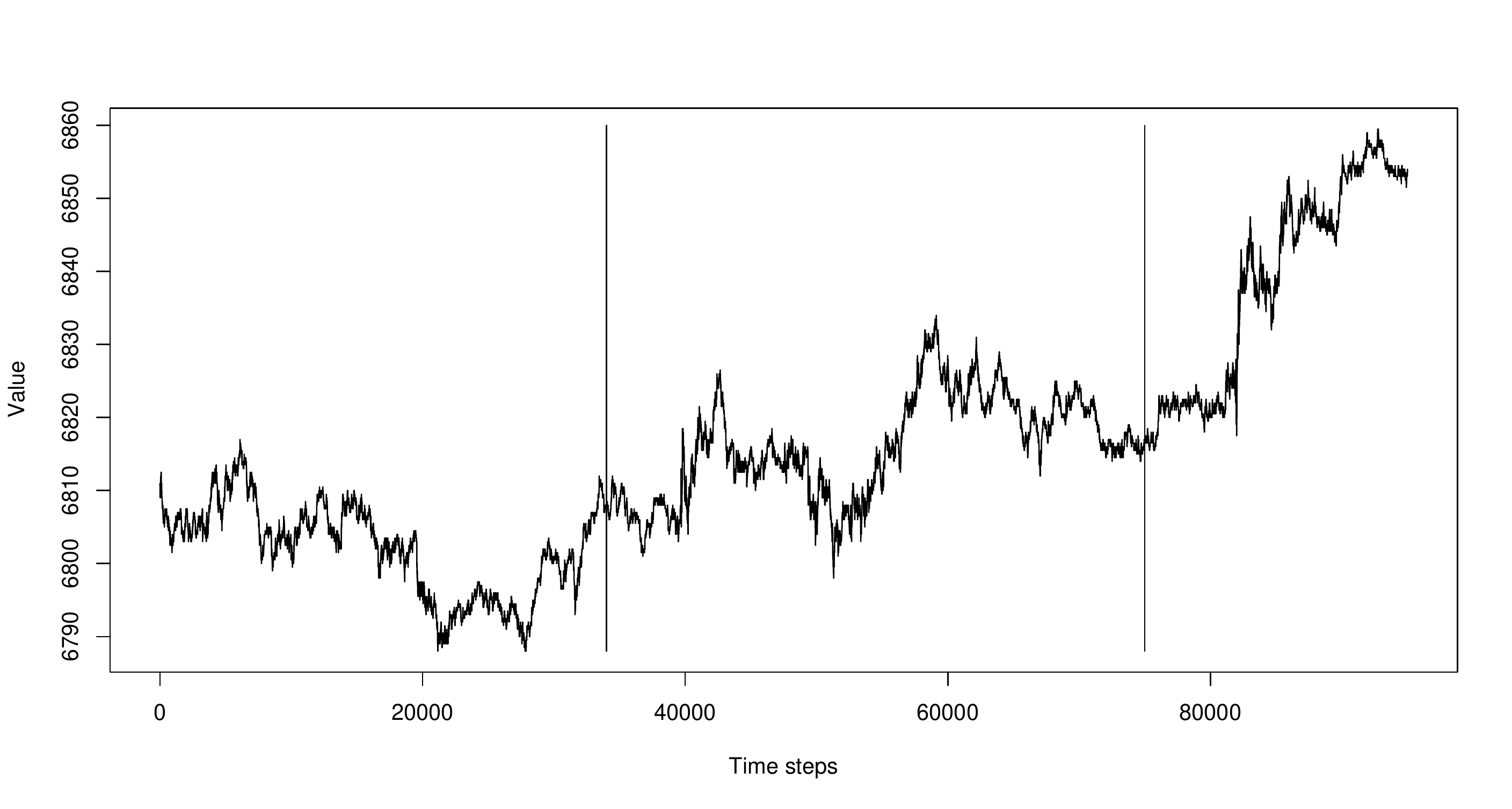}
\caption{Boundaries found for 3 segments.}
\label{fig160}
\end{figure*}

Reading off the 3-cluster memberships from Figure \ref{fig140} 
gives for the signal actually used (with a very initial segment 
and a very final segment deleted): cluster 1 corresponds to signal 
values 1000 to 33999 (points 1 to 33 in Figure \ref{fig140}); 
cluster 2 corresponds to signal values 34000 to 74999 (points 34 to 74 
in Figure \ref{fig140}); 
and cluster 3 corresponds to signal values 75000 to 86999 (points 
75 to 86 in Figure \ref{fig140}).  
This allows us to segment the original time series: see Figure \ref{fig160}.
(The clustering of the 85-dimensional embedding differs minimally.  
Segments are: points 1 to 32; 33 to 73; and 74 to 86.)

To summarize what has been done:

\begin{enumerate}
\item the segmentation is initially guided by the peak-finding in the 
histogram of distances
\item with high dimensionality we expect simple structure in a low
dimensional mapping provided by principal coordinates analysis
\item which we use as input to a sequence-constrained clustering 
method in order to determine the clusters
\item which can then be displayed on the original data. 
\end{enumerate}

In this case, the clusters are defined using a complete link criterion, 
implying that these three clusters
are determined by minimizing their maximum internal pairwise distance.
This provides a strong measure of signal volatility as an  
explanation for the clusters, in addition to their average value.  

\section{Conclusions}


One  interesting conclusion on this work follows.
Traditionally, clustering algorithms have generally been considered as
distance-based or model-based.
The former is exemplified by agglomerative hierarchical clustering,
or k-means partitioning.
The latter is exemplified by Gaussian mixture modeling.  (One
motivation for model-based clustering is the computational difficulty, 
in general, of taking account of all pairwise distances.)
The approach described in this work is both distance-based and
model-based.

 
What we have observed in all of 
this work is that in the limit of high dimensionality 
a scalar product space becomes ultrametric.  
It has been our aim in this work to link observed data with an ultrametric
topology for such data.  The traditional approach in data analysis, of course, 
is to impose structure on the data.  This is done, for example, by using 
some agglomerative hierarchical clustering algorithm.  We can always 
do this (modulo distance or other ties in the data).  Then we can assess
the degree of fit of such a (tree or other) structure to our data.  
For our purposes, here, this is unsatisfactory.  

\begin{itemize}
\item Firstly, our aim was to show 
that ultrametricity can be naturally present in our data, globally or 
locally.  We did not want any ``measuring tool'' such as an 
agglomerative hierarchical clustering algorithm to overly influence 
this finding.  (Unfortunately Rammal et al., 1986, suffers from precisely this 
unhelpful influence of the ``measuring tool'' of the subdominant 
ultrametric.  In other respects, Rammal et al., 1986, is a seminal paper.) 

\item
Secondly, let us assume that we did use hierarchical clustering, and then 
based our discussion around the goodness of fit.  This again is a traditional
approach used in data analysis, and in statistical data modeling.  But such 
a discussion would have been unnecessary and futile.  For, after all, if 
we have ultrametric properties in our data then many of the widely used
hierarchical clustering algorithms will give precisely the same outcome, 
and furthermore the fit is by definition optimal.   (Our point here is 
that if $\min \{ d_{ik} | i \in q, k \not\in q, k\neq q \} =
\max \{ d_{ik} | i \in q, k \not\in q, k\neq q \}$ for cluster $q$, at 
all agglomerations, then single linkage and complete linkage are identical.) 

\end{itemize}

We have described an application of this work to very high frequency 
signal processing.  The twin objectives are signal segmentation, and model 
identification.  We have noted that a considerable amount of this work is 
model-based: we require assumptions (on clusters, and on model(s)) for 
identifiability.  

Motivation for this work includes the availability of very high frequency 
data streams in various fields (physics, 
engineering, finance, meteorology, 
bio-engineering,  and bio-medicine).  By using a very large embedding 
dimensionality, we are approaching the data analysis on a very gross 
scale, and hence furnishing a particular type of multiresolution analysis.  
That this is worthwhile has been shown in our case studies.  

\section*{References}

AGGARWAL, C.C., HINNEBURG, A. and KEIM, D.A. (2001). 
``On the
Surprising Behavior of Distance Metrics in High Dimensional Spaces'',
{\em Proceedings of the 8th International Conference on Database Theory},
pp.\ 420--434, January 04-06.

\medskip
\noindent
AHN, J., MARRON, J.S., MULLER, K.E. and CHI, Y.-Y. (2007).  
``The High Dimension,
Low Sample Size Geometric Representation Holds under Mild Conditions'',
{\em Biometrika}, 94, 760--766.

\medskip
\noindent
AHN, J. and MARRON, J.S. (2005).  ``Maximal Data Piling in Discrimination'', 
{\em Biometrika}, submitted;  and 
``The Direction of Maximal Data Piling in High Dimensional Space''.  



\medskip
\noindent
BELLMAN, R. (1961). {\em Adaptive Control Processes: A Guided Tour}, 
Princeton University Press.

\medskip
\noindent
B\'ENASS\'ENI, J., BENNANI DOSSE, M. and JOLY, S. (2007).  
On a General Transformation Making a 
Dissimilarity Matrix Euclidean, {\em Journal of Classification}, 24, 33--51.

\medskip
\noindent
BENZ\'ECRI, J.P. (1979).  
{\em L'Analyse des Donn\'ees, Tome I Taxinomie}, 
{\em Tome II Correspondances}, 2nd ed., Dunod, Paris.

\medskip
\noindent
BREUEL, T.M. (2007). 
``A Note on Approximate Nearest Neighbor Methods'',
http://arxiv.org/pdf/cs/0703101


\medskip
\noindent
CAILLIEZ, F. and PAG\`ES, J.P. (1976). 
{\em Introduction \`a l'Analyse de
  Donn\'ees}, SMASH (Soci\'et\'e de Math\'ematiques Appliqu\'ees et de
  Sciences Humaines), Paris.

\medskip
\noindent
CAILLIEZ, F. (1983). 
``The Analytical Solution of the Additive Constant Problem'', {\em Psychometrika}, 48, 305--308.


\medskip
\noindent
CH\'AVEZ, E., NAVARRO, G., BAEZA-YATES, R. and MARROQU\'IN, J.L. (2001).
``Proximity Searching in Metric Spaces'', 
{\em ACM Computing Surveys}, 33,
  273--321.


\medskip
\noindent
CRITCHLEY, F. and HEISER, W. (1988),
``Hierarchical trees can be perfectly scaled in one dimension''
{\em Journal of Classification}, 5, 5--20.

\medskip
\noindent
DE SOETE, G. (1986). 
``A Least Squares Algorithm for Fitting an Ultrametric Tree 
to a Dissimilarity Matrix'',
{\em Pattern Recognition Letters}, 2, 133--137.


\medskip
\noindent
DONOHO, D.L. and TANNER, J. (2005).
``Neighborliness of Randomly-Projected Simplices in High Dimensions'',
{\em Proceedings of the National Academy of Sciences}, 102, 9452--9457.







\medskip
\noindent
HALL, P., MARRON, J.S. and NEEMAN, A. (2005). 
``Geometric Representation of High Dimension Low Sample Size Data'',
{\em Journal of the Royal Statistical Society B}, 67, 427--444.

\medskip
\noindent
HEISER, W.J. (2004).  ``Geometric Representation of Association between
Categories'', {\em Psychometrika}, 69, 513--545.

\medskip
\noindent
HINNEBURG, A., AGGARWAL, C. and KEIM, D. (2000).  
``What is the Nearest Neighbor
in High Dimensional Spaces?'',
{\em VLDB 2000, Proceedings of 26th International Conference on Very
               Large Data Bases}, September 10-14, 2000, Cairo, Egypt,
Morgan Kaufmann, pp.\ 506--515.

\medskip
\noindent
HORNIK, K. (2005). 
``A CLUE for CLUster Ensembles'', 
{\em Journal of Statistical Software}, 14 (12).

\medskip
\noindent
KASS, R.E. and RAFTERY, A.E. (1995). 
``Bayes Factors and Model Uncertainty'',
{\em Journal of the American Statistical Association}, 90, 773--795. 

\medskip
\noindent
KHRENNIKOV, A. (1997).
{\em Non-Archimedean Analysis: Quantum Paradoxes, Dynamical
Systems and Biological Models}, Kluwer.

\medskip
\noindent
LERMAN, I.C. (1981). 
{\em Classification et Analyse Ordinale des Donn\'ees}, 
Paris, Dunod.


\medskip
\noindent
MURTAGH, F. (1985). 
{\em Multidimensional Clustering Algorithms}, Physica-Verlag.


\medskip
\noindent
MURTAGH, F. (2004). 
``On Ultrametricity, Data Coding, and Computation'', 
{\em Journal of Classification}, 21, 167--184.

\medskip
\noindent
MURTAGH, F. (2005). 
``Identifying the Ultrametricity of Time Series'', 
{\em European Physical Journal B}, 43, 573--579.

\medskip
\noindent
MURTAGH, F. (2007). 
``A Note on Local Ultrametricity in Text'', \\
http://arxiv.org/pdf/cs.CL/0701181
 
\medskip
\noindent
MURTAGH, F. (2005).  
{\em Correspondence Analysis and Data Coding with R and Java},
Chapman \& Hall/CRC.

\medskip
\noindent
MURTAGH, F. (2006). 
``From Data to the Physics using Ultrametrics: New Results
in High Dimensional Data Analysis'', 
in A.Yu. Khrennikov, Z. Raki\'c and I.V.
Volovich, Eds., {\em p-Adic Mathematical Physics}, American Institute of Physics
Conf.\ Proc.\ Vol.\ 826,
151--161.

\medskip
\noindent
MURTAGH, F., DOWNS, G. and CONTRERAS, P. (2008).
``Hierarchical Clustering of Massive, High Dimensional 
Data Sets by Exploiting Ultrametric Embedding'', 
{\em SIAM Journal on Scientific Computing}, 30, 707--730. 

\medskip
\noindent
MURTAGH, F. and STARCK, J.L. (2003). 
``Quantization from Bayes Factors with 
Application to Multilevel Thresholding'', {\em Pattern Recognition Letters}, 
24, 2001--2007.



\medskip
\noindent
NEUWIRTH, E. and REISINGER, L. (1982). 
``Dissimilarity and Distance Coefficients
in Automation-Supported Thesauri'',  {\em Information Systems}, 
7, 47--52.



\medskip
\noindent
RAMMAL, R., ANGLES D'AURIAC, J.C. and DOUCOT, B. (1985). 
``On the Degree of Ultrametricity'', 
{\em Le Journal de Physique -- Lettres}, 46, L-945--L-952.

\medskip
\noindent
RAMMAL, R., TOULOUSE, G. and VIRASORO, M.A. (1986). 
``Ultrametricity for Physicists'', 
{\em Reviews of Modern Physics}, 58, 765--788.

\medskip
\noindent
ROHLF, F.J. and FISHER, D.R. (1968). 
``Tests for Hierarchical Structure in Random
Data Sets'', {\em Systematic Zoology}, 17, 407--412.


\medskip
\noindent
SCHWARZ, G. (1978).
``Estimating the Dimension of a Model'', {\em Annals of Statistics}, 6, 
461--464.

\medskip
\noindent
TORGERSON, W.S. (1958). 
{\em Theory and Methods of Scaling}, Wiley.

\medskip
\noindent
TREVES, A. (1997). 
``On the Perceptual Structure of Face Space'',
{\em BioSystems}, 40, 189--196.




\end{document}